\pdfoutput=1
\documentclass[5p,twocolumn]{elsarticle}

\biboptions{sort&compress}

\usepackage{amsmath,amssymb,graphicx,booktabs,multirow,array,microtype,placeins,adjustbox,balance}
\usepackage[labelformat=parens,labelfont=bf,font=footnotesize,skip=2.6pt]{subcaption}
\usepackage[hyphens]{url}

\usepackage{newtxtext,newtxmath}
\usepackage[colorlinks=true,citecolor=blue,linkcolor=blue,urlcolor=blue]{hyperref}
\makeatletter
\def\ps@pprintTitle{%
  \let\@oddhead\@empty
  \let\@evenhead\@empty
  \def\@oddfoot{\reset@font\hfil\thepage\hfil}%
  \let\@evenfoot\@oddfoot
}
\makeatother

\newcommand{\AuthorName}{Pavel Chuklanov}
\newcommand{\AuthorEmail}{chuklanov.pavel@gmail.com}
\newcommand{\AuthorOrg}{Moscow Institute of Physics and Technology}
\newcommand{\AuthorStreet}{Institutsky lane 9}
\newcommand{\AuthorCity}{Dolgoprudny, Moscow region}
\newcommand{\AuthorPostcode}{141700}
\newcommand{\AuthorCountry}{Russia}

\begin{document}

\begin{frontmatter}

\title{Empirical De-Quantization and Norm-Controlled Spectral Audit of Quantum Cognition Machine Learning in an Emerging Equity Market}

\author[1]{\AuthorName\corref{cor1}}
\ead{\AuthorEmail}
\cortext[cor1]{Corresponding author.}
\affiliation[1]{organization={\AuthorOrg},
                addressline={\AuthorStreet},
                city={\AuthorCity},
                postcode={\AuthorPostcode},
                country={\AuthorCountry}}

\begin{abstract}
Quantum-inspired machine learning in econophysics describes states of financial markets through matrix Hamiltonians, using ground-state expectations to predict returns and spectral gaps to identify crises. Although early work reports encouraging advances, it remains unclear whether these stem from genuine noncommutative quantum geometry or from conventional classical regularization and scale effects. We test this in a controlled audit of matrix-form Quantum Cognition Machine Learning (QCML) and variational quantum circuits on the Moscow Exchange (2020--2025), combining daily equity data with causally lagged macro news under realistic trading costs, short-selling bans, and abrupt shifts in market regimes. By using capacity-matched operator swaps---which progressively eliminate complex Berry phases and matrix noncommutativity down to a commuting classical Gibbs radial-basis limit under scale-invariant regularization---as well as testing spectral indicators against radial-norm null hypotheses and untrained GUE random matrices, we isolate the contribution of each quantum element. In out-of-sample data, neither complex phases, noncommutativity, nor circuit entanglement improves the ranking of returns relative to classical controls. Once feature magnitudes are normalized, apparent crisis detection from raw spectral gaps drops below chance levels. These results show that on noisy equity panels, matrix-form QCML acts as an adaptive spectrally bounded classical smoothing kernel: it reliably integrates lagged news and curbs extreme crisis forecasts without passing economic multiple-testing thresholds, while our audit provides operator-swap and random-matrix controls for evaluating future quantum-inspired financial models.
\end{abstract}

\begin{keyword}
Econophysics \sep Statistical mechanics of machine learning \sep Non-commutative geometry \sep Random matrix theory \sep Financial regime shifts \sep Moscow Exchange
\end{keyword}

\end{frontmatter}

\section{Introduction}\label{sec:intro}

From density operator-based risk indices to Hamiltonian embeddings, statistical-mechanical and operator-based models of markets aim to separate true structure from finite-sample noise \cite{laloux_1999,plerou_2002,gong_2025,candelori_2025,abanov_2025}. When daily stock panels combine weak signal with short histories and abrupt regime changes, standard regressions, boosted trees, and neural networks are susceptible to variance inflation and out-of-sample degradation \cite{campbell_2008,gu_2020,grinsztajn_2022,samson_2024,shen_2026}. Quantum Cognition Machine Learning (QCML) represents each standardized feature vector $\mathbf{x}_{i,t} \in \mathbb{R}^p$ as the ground state $\lvert\psi_0(\mathbf{x}_{i,t})\rangle$ of a learned $d \times d$ Hermitian error Hamiltonian $H(\mathbf{x}_{i,t})$ \cite{berenstein_2012,ishiki_2015,steinacker_2021,candelori_2025,abanov_2025}, with a single target observable specifying the bounded return forecast $\hat{y}_{i,t}$. The ground state energy $E_0(\mathbf{x}_{i,t})$ and the first spectral gap $\Delta E(\mathbf{x}_{i,t})$ serve as built-in geometric diagnostics. Emerging markets pose a severe stress test for both properties. Over 2020--2025, the Moscow Exchange (MOEX) experienced heavy-tailed volatility, a trading halt in February--March 2022, a 47-session ban on short selling by the Central Bank of Russia (CBR) and policy rate fluctuations from $4.25\%$ to $21.00\%$ \cite{kraevskiy_2024,teplova_2022,khubiev_2025}. Validation of multimodal market and news models over this period also requires strictly lagged non-LLM text indicators to prevent pre-trained language models from memorizing future data \cite{lopez_lira_2025,kirtac_2024,halder_2022}.

Despite the rise of quantum and quantum-inspired finance \cite{orus_2019,herman_2023,musaelian_2024}, four gaps obscure the source of the claimed gains: non-commutative quantum geometry or classical regularization. First, recent operator regime-detection studies link crisis separation to risk indices based on density operators \cite{gong_2025} and to QCML geometric observables and spectral gaps \cite{hammond_2026}; in the error-Hamiltonian QCML setting this happens despite the known finite-dimensional matrix-geometry artifacts \cite{felder_2024} and without controlling for the squared feature norm $\lVert\mathbf{x}_{i,t}\rVert_2^2$ or testing untrained random matrices. The second gap is that applied finance work on QCML relies on static partitions without tree baselines, transaction costs and commuting operator controls \cite{samson_2024}, or is limited to single cross-sections of bond pricing \cite{rosaler_2025}, pre-cost firm similarity \cite{samson_2025}, or non-financial manifolds \cite{candelori_2025,caro_2025}. At the same time, theoretical work shows that quantum-like models of sequential decision making allow for classical instrument representations \cite{tull_2026}, whereas discussions around Hamiltonian embeddings stress comparable classical baselines \cite{molteni_2026,abanov_2026}. The third gap is that variational quantum circuits (VQCs) \cite{mitarai_2018,cerezo_2021} and quantum kernel methods \cite{havlicek_2019,schuld_killoran_2019,huang_2021,schuld_2021} encounter barren plateaus \cite{mcclean_2018,cerezo_2021_bp,larocca_2025,cerezo_2023} and exponential concentration \cite{thanasilp_2024}; however, many financial studies of QML omit strong tree benchmarks, trading frictions and multiple testing adjustments \cite{emmanoulopoulos_2022,kobayashi_2023,xu_2024,choudhary_2026}. Fourth, after 2024, systematic audits in machine learning and finance consistently reveal benchmark selection bias \cite{bowles_2024}, equal performance without quantum entanglement \cite{schnabel_2025}, the dominance of classical methods on time series \cite{jones_2024,fellner_2026,ahmad_2026} and the disappearance of gains in stock forecasting under strict point-in-time protocols \cite{shen_2026}.

We consider matrix-form QCML ($d \in \{2,4,8,16\}$) and noise-free 3-qubit VQCs strictly as exact classical algorithms, with no claims of quantum hardware acceleration \cite{li_2026,thakkar_2024}. Our empirical study relies on a split-adjusted $1{,}251$-day panel of MOEX containing $37{,}048$ unfiltered out-of-sample stock-day observations for $759$ trading sessions and six expanding walk-forward folds over a five-day horizon $h = 5$. For market-only and market-with-news inputs, we compare QCML with parameter-free factors, linear regressions, kernels, boosted trees, neural networks, VQCs, and commuting operator substitutions. Three main conclusions stand out:
\begin{enumerate}
    \item \emph{Empirical de-quantization of return ranking.} With dimension-invariant $1/d$ weight decay and equal parameter budgets, neither complex phases, nor non-commutativity of operators, nor circuit entanglement improves out-of-sample ranking compared to real-symmetric QCML, a classical commuting radial-basis Gibbs swap or separable circuits. Lagged news improves both QCML and its classical commuting swap. Energy anchoring acts primarily as a variance stabilizer, cutting cross-seed dispersion in half.
    \item \emph{Crisis detection from the spectral gap as an artifact of the radial norm.} The apparent detection of crises from raw $\Delta E(\mathbf{x}_{i,t})$ and raw Granger causality is reproduced by untrained random matrices from a Gaussian Unitary Ensemble (GUE), since the eigenvalues mechanically scale with $\lVert\mathbf{x}_{i,t}\rVert_2$. After normalizing the feature magnitude, the spectral separation of crises falls below chance, and Bonferroni-significant forward Granger causality disappears.
    \item \emph{Portfolio frictions versus clamping crisis forecasts.} With a one-day execution lag, short selling bans and $15\,\text{bps}$ one-way costs, the leading models give positive active information ratios over the equal-weight market, but none pass the multiple testing threshold for the Deflated Sharpe Ratio ($\mathrm{DSR} > 0.95$) \cite{bailey_2014,harvey_2016}. In contrast, the eigenvalue bounds constrain extreme forecast errors during the February--March 2022 halt relative to unconstrained tree and neural models.
\end{enumerate}

Section~\ref{sec:methods} introduces the Hamiltonian for QCML, its commuting classical limit and spectral scaling in the radial norm. Analytical derivations of gradients and a glossary of notation are contained in \ref{app:derivations}. A description of the MOEX dataset, lagged news construction, and walk-forward evaluation protocol is provided in Section~\ref{sec:data}, while Section~\ref{sec:results} presents audits of forecasting, spectral gaps, and portfolios, supplemented by synthetic positive controls, multi-output forecasts, and robustness checks in the Supplementary Material (Sections~S1--S3). Finally, Section~\ref{sec:discussion} discusses implications, limitations, and conclusions.

\section{Matrix-Form QCML, Operator-Swap Controls, and Spectral Scaling}\label{sec:methods}

\subsection{Error Hamiltonian, Commuting Gibbs Limit, and Regularized Training}\label{subsec:qcml_core}

\begin{figure*}[t]
  \centering
  \begin{subfigure}[t]{0.42\textwidth}
    \includegraphics[width=\linewidth]{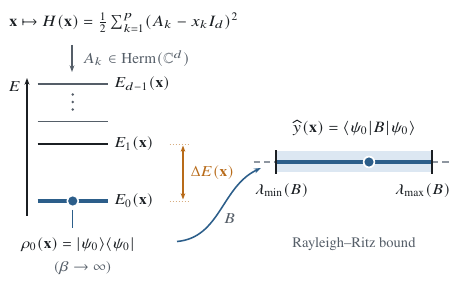}
    \caption{}
  \end{subfigure}\hfill
  \begin{subfigure}[t]{0.56\textwidth}
    \includegraphics[width=\linewidth]{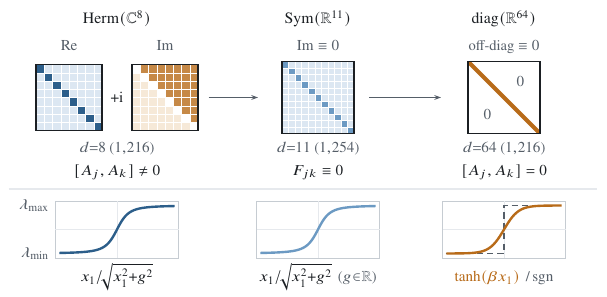}
    \caption{}
  \end{subfigure}

  \vspace{2.8pt}
  \begin{subfigure}[t]{0.99\textwidth}
    \includegraphics[width=\linewidth]{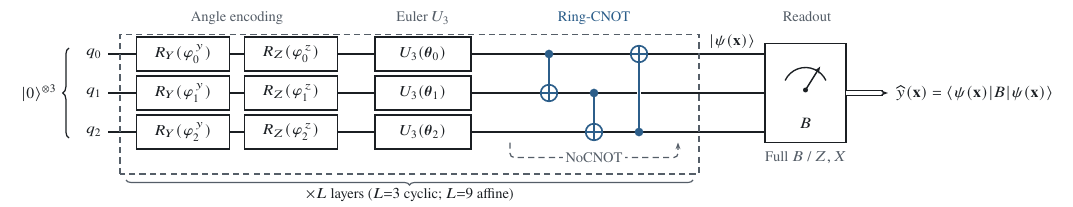}
    \caption{}
  \end{subfigure}
  \caption{Model architectures and operator-swap controls. (\textbf{a})~Matrix-form QCML sets the error Hamiltonian $H(\mathbf{x})$, spectral gap $\Delta E(\mathbf{x})$, ground-state projector $\rho_0(\mathbf{x}) = \lvert\psi_0\rangle\langle\psi_0\rvert$ and bounded readout $\hat{y}(\mathbf{x}) \in [\lambda_{\min}(B),\lambda_{\max}(B)]$. (\textbf{b})~On the market-plus-news specification (\texttt{M+N}, $1{,}216$--$1{,}254$ parameters), the equal-capacity ladder of operator swaps tracks non-commutativity ($\kappa_{\text{nc}}$) and Berry curvature ($F_{jk}$) through complex Hermitian ($d=8$), phase-free real-symmetric ($d=11$, $F_{jk}\equiv 0$) and commuting diagonal ($d=64$, $\kappa_{\text{nc}}\equiv 0$) matrices, and the lower strip compares the exact one-dimensional readout slices $\hat{y}(x_1)$ from Eqs.~\eqref{eq:hamiltonian}--\eqref{eq:gibbs_readout}, contrasting non-commutative smoothing ($x_1/\sqrt{x_1^2+g^2}$, with off-diagonal coupling $g > 0$) with finite-temperature Gibbs ($\tanh(\beta x_1)$ at finite $\beta$) and the Voronoi step ($\operatorname{sgn}(x_1)$ at $\beta=\infty$). (\textbf{c})~Three-qubit data-reuploading VQC compares cyclic ($L=3$, $200$ parameters) and affine ($L=9$, $1{,}118$ parameters) encodings with switchable ring-CNOT against separable \texttt{NoCNOT} layers with full Hermitian or local Pauli-$Z/X$ readout.}
  \label{fig:circuits}
\end{figure*}

Matrix-form QCML represents $p$ standardized input variables $\mathbf{x}_{i,t} \in \mathbb{R}^p$ (with coordinates $x_{i,t,k}$) via $p$ learnable $d \times d$ Hermitian matrices $\mathcal{A} = \{A_k\}_{k=1}^{p}$ \cite{berenstein_2012,ishiki_2015,steinacker_2021,felder_2024} along with a Hermitian target matrix $B$ for return readout \cite{samson_2024,rosaler_2025}. For each stock $i$ and day $t$ the model forms a positive semidefinite error Hamiltonian $H(\mathbf{x}_{i,t})$. The input vector is then mapped to its normalized ground-state eigenvector $\lvert\psi_0(\mathbf{x}_{i,t})\rangle \in \mathbb{C}^d$ (Fig.~\ref{fig:circuits}a):
\begin{equation}\label{eq:hamiltonian}
\begin{aligned}
H(\mathbf{x}_{i,t}) &= \frac{1}{2}\sum_{k=1}^{p}\bigl(A_k - x_{i,t,k} I_d\bigr)^2 \\
&= \tilde{H}(\mathbf{x}_{i,t}) + \frac{1}{2}\lVert\mathbf{x}_{i,t}\rVert_2^2 I_d, \\
H(\mathbf{x}_{i,t})\lvert\psi_0(\mathbf{x}_{i,t})\rangle &= E_0(\mathbf{x}_{i,t})\lvert\psi_0(\mathbf{x}_{i,t})\rangle,
\end{aligned}
\end{equation}
where $\tilde{H}(\mathbf{x}_{i,t}) = C_0(\mathcal{A}) - \sum_{k=1}^{p} x_{i,t,k} A_k$, $C_0(\mathcal{A}) = \frac{1}{2}\sum_{k=1}^{p} A_k^2$, and eigenvalues are ordered $0 \le E_0(\mathbf{x}) \le E_1(\mathbf{x}) \le \dots \le E_{d-1}(\mathbf{x})$.

Equation~\eqref{eq:hamiltonian} allows direct statistical interpretation \cite{candelori_2025}. For any unit state $\lvert\psi\rangle$, where the expected coordinate is written as $\bar{x}_k(\psi) = \langle\psi\vert A_k\vert\psi\rangle$ and the total quantum variance is $\sigma^2(\psi) = \sum_{k=1}^{p}\bigl(\langle\psi\vert A_k^2\vert\psi\rangle - \bar{x}_k(\psi)^2\bigr)$, the algebraic expansion gives $E_0(\mathbf{x}) = \frac{1}{2}\lVert\bar{\mathbf{x}}(\psi_0) - \mathbf{x}\rVert_2^2 + \frac{1}{2}\sigma^2(\psi_0)$. Therefore, the minimization of $H(\mathbf{x})$ balances the classical reconstruction error against the irreducible variance over non-commuting matrices and places each observation on a learned state manifold \cite{candelori_2025,hammond_2026}.

To directly connect the zero-temperature ground-state QCML model with classical kernel smoothing, we write the return forecast through a finite-temperature Gibbs density matrix $\rho_\beta(\mathbf{x})$ at inverse temperature $\beta > 0$:
\begin{equation}\label{eq:gibbs_readout}
\begin{aligned}
\hat{y}_{i,t} &= \operatorname{Tr}\!\bigl(\rho_\beta(\mathbf{x}_{i,t})\, B\bigr) \in \bigl[\lambda_{\min}(B),\,\lambda_{\max}(B)\bigr], \\
\rho_\beta(\mathbf{x}) &= \frac{e^{-\beta H(\mathbf{x})}}{\operatorname{Tr}\!\bigl(e^{-\beta H(\mathbf{x})}\bigr)}.
\end{aligned}
\end{equation}
In the limit $\beta \to \infty$, the matrix $\rho_\beta(\mathbf{x})$ reduces to the pure ground state projector $\rho_0(\mathbf{x}) = \lvert\psi_0(\mathbf{x})\rangle\langle\psi_0(\mathbf{x})\rvert$, which implies $\hat{y}_{i,t} = \langle\psi_0(\mathbf{x}_{i,t})\vert B\vert\psi_0(\mathbf{x}_{i,t})\rangle$. Even when extreme crisis shocks drive $\lVert\mathbf{x}_{i,t}\rVert_2$ far beyond the training range, according to the Rayleigh--Ritz inequality, each forecast in Eq.~\eqref{eq:gibbs_readout} remains bounded between the extreme eigenvalues ($[\lambda_{\min}(B), \lambda_{\max}(B)]$).

When the matrix algebra $(A_k, B)$ is constrained at a matched number of parameters, a three-step de-quantization ladder arises (Fig.~\ref{fig:circuits}b). At the first stage, complex Hermitian matrices ($d = 8$, $1{,}216$ parameters on $p = 18$ inputs) preserve complex phases and non-commutativity ($[A_j, A_k] \neq 0$), and at the second stage, real-symmetric matrices ($d = 11$, $1{,}254$ parameters) discard complex phases and Berry curvature $F_{jk}$ ($F_{jk} \equiv 0$; \ref{app:gradients}), while maintaining non-commutativity. Commuting diagonal matrices $A_k = \operatorname{diag}(a_{k,0}, \dots, a_{k,d-1})$ ($d = 64$, $1{,}216$ parameters) at the third stage completely remove non-commutativity ($[A_j, A_k] = 0$). Diagonal QCML at $\beta = \infty$ degenerates into a piecewise constant choice of the nearest prototype with zero local gradients, while at finite $\beta$ Eq.~\eqref{eq:gibbs_readout} becomes a smooth classical radial basis kernel (Gibbs--RBF) over $d$ learned prototypes with softmax weights $w_\alpha(\mathbf{x}) \propto \exp\bigl(-\frac{\beta}{2}\sum_{k=1}^{p}(a_{k,\alpha} - x_k)^2\bigr)$ \cite{candelori_2025,abanov_2025}. In parallel, we test noiseless 3-qubit data-reuploading VQCs \cite{perez_salinas_2020,schuld_sweke_2021} (cyclic $L = 3$ and affine $L = 9$ encodings), in which the ring-CNOT gates are turned on or off (\texttt{NoCNOT}) to separate circuit entanglement from the readout capacity (Fig.~\ref{fig:circuits}c) \cite{kobayashi_2023,bowles_2024,schnabel_2025}.

For a fixed scale of eigenvalues, the squared Frobenius norm $\lVert A_k\rVert_F^2$ increases in proportion to $d$. Standard weight decay would therefore excessively penalize the $d = 64$ diagonal control (by $8\times$) relative to the $d = 8$ Hermitian QCML, and in the training objective we normalize the Frobenius regularization by $1/d$:
\begin{equation}\label{eq:loss}
\begin{aligned}
\mathcal{L}(\mathcal{A}, B) &= \frac{1}{N}\sum_{(i,t)}\Bigl[\bigl(\hat{y}_{i,t} - y_{i,t}\bigr)^2 + \lambda_E\, E_0(\mathbf{x}_{i,t})\Bigr] \\
&\quad + \frac{\lambda_{\text{reg}}}{d}\sum_{k=1}^{p}\lVert A_k\rVert_F^2 + \frac{\lambda_{\text{reg}, B}}{d}\lVert B\rVert_F^2.
\end{aligned}
\end{equation}
Varying $\lambda_E$ covers energy-anchored ($\lambda_E = 0.05$) and unanchored ($\lambda_E = 0$) supervised learning as well as unsupervised learning ($\lambda_E = 1.0$); for spectral tests of market regimes, we also include the variance-weighted unsupervised loss $\mathcal{L}_{\text{Cand}}(\mathcal{A}) = \frac{1}{N}\sum_{(i,t)}\bigl(\lVert\bar{\mathbf{x}}(\psi_0) - \mathbf{x}_{i,t}\rVert_2^2 + w_{\text{var}}\sigma^2(\psi_0)\bigr)$ with $w_{\text{var}} = 0.1$ and the same $1/d$ penalty \cite{candelori_2025}. Forward evaluation diagonalizes $\tilde{H}(\mathbf{x})$ in $O(d^3)$ time per observation. Exact gradients are computed via regularized perturbation theory (\ref{app:derivations}).

\subsection{Radial Norm Confound and Non-Commutativity Index}\label{subsec:norm_scaling}

\begin{figure*}[t]
  \centering
  \begin{subfigure}[t]{0.32\textwidth}
    \includegraphics[width=\linewidth]{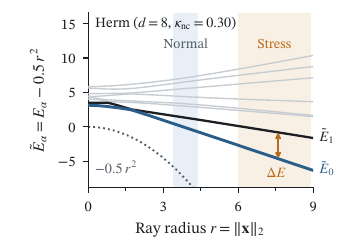}
    \caption{}
  \end{subfigure}\hfill
  \begin{subfigure}[t]{0.32\textwidth}
    \includegraphics[width=\linewidth]{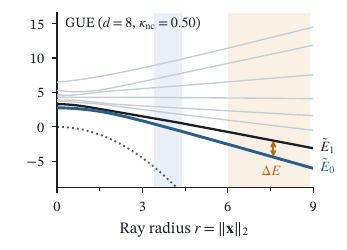}
    \caption{}
  \end{subfigure}\hfill
  \begin{subfigure}[t]{0.32\textwidth}
    \includegraphics[width=\linewidth]{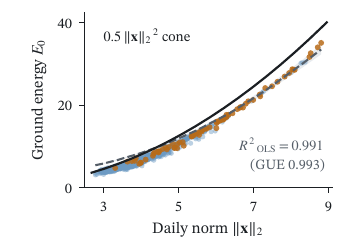}
    \caption{}
  \end{subfigure}

  \vspace{2.8pt}
  \begin{subfigure}[t]{0.485\textwidth}
    \includegraphics[width=\linewidth]{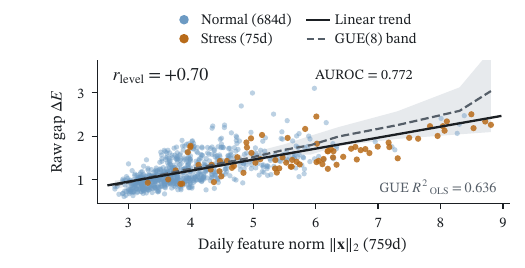}
    \caption{}
  \end{subfigure}\hfill
  \begin{subfigure}[t]{0.485\textwidth}
    \includegraphics[width=\linewidth]{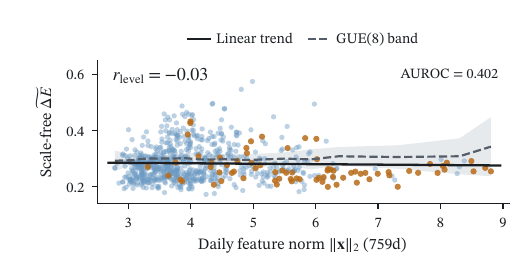}
    \caption{}
  \end{subfigure}
  \caption{Radial-norm spectral scaling and random-matrix null. (\textbf{a}, \textbf{b})~Along a fixed feature ray, exact positive-semidefinite shifted Stark level fans $\tilde{E}_\alpha(r) = E_\alpha(r) - \frac{1}{2}r^2$ ($r = \lVert\mathbf{x}\rVert_2$, dotted $-\frac{1}{2}r^2$ floor enforcing $E_0 \ge 0$) for anchored Hermitian matrices ($d = 8$, $\kappa_{\text{nc}} = 0.30$) and an untrained random GUE draw ($d = 8$, $\kappa_{\text{nc}} = 0.50$) show avoided crossings at low $r$ and linear opening of the $\Delta E$ gap in the normal and stress MOEX norm bands. (\textbf{c}--\textbf{e})~When the supervised daily spectrum of \texttt{M+N} for $759$ out-of-sample days ($684$ normal, $75$ stress) is superimposed on ten untrained GUE($8$) draws ($\pm 1\text{ SD}$), the ground-state energy $E_0$ follows $\frac{1}{2}\lVert\mathbf{x}\rVert_2^2$ ($R^2_{\text{OLS}} = 0.991$, GUE $0.993$), the raw gap $\Delta E$ grows with $\lVert\mathbf{x}\rVert_2$ (quadratic GUE $R^2_{\text{OLS}} = 0.636$; daily $r_{\text{level}} = +0.70$, crisis area under the ROC curve (AUROC) $0.772$), and scale-free $\widetilde{\Delta E}$ eliminates the norm trend ($r_{\text{level}} = -0.03$, AUROC $0.402$).}
  \label{fig:algo_geom}
\end{figure*}

When $\mathbf{x} = \lVert\mathbf{x}\rVert_2 \hat{\mathbf{x}}$ is written through the feature amplitude $\lVert\mathbf{x}\rVert_2$ and the direction on the unit sphere $\hat{\mathbf{x}} \in S^{p-1}$, a mechanical scaling confound appears in Eq.~\eqref{eq:hamiltonian}: $H(\mathbf{x}) = \frac{1}{2}\lVert\mathbf{x}\rVert_2^2 I_d - \lVert\mathbf{x}\rVert_2 \sum_{k=1}^{p}\hat{x}_k A_k + C_0(\mathcal{A})$. Spikes in stock returns and volatility during market stress inflate $\lVert\mathbf{x}\rVert_2$. According to Weyl's perturbation inequality, $E_0(\mathbf{x})$ then grows quadratically as $\frac{1}{2}\lVert\mathbf{x}\rVert_2^2 + O(\lVert\mathbf{x}\rVert_2)$, and the raw spectral gap $\Delta E(\mathbf{x}) = E_1(\mathbf{x}) - E_0(\mathbf{x})$ grows linearly with $\lVert\mathbf{x}\rVert_2$ for \emph{any} fixed matrices, including untrained random GUE matrices (Fig.~\ref{fig:algo_geom}a--e; Table~\ref{tab:spectral_audit}). To separate the genuine directional geometry of market regimes from feature amplitude, we calculate the scale-free spectral gap
\begin{equation}\label{eq:gap_scalefree}
\widetilde{\Delta E}(\mathbf{x}) = \frac{\Delta E(\mathbf{x})}{\sqrt{1 + \lVert\mathbf{x}\rVert_2^2}},
\qquad
\Delta E(\mathbf{x}) = E_1(\mathbf{x}) - E_0(\mathbf{x}),
\end{equation}
along with unit-sphere evaluations $E_0^{\text{sphere}}(\mathbf{x}) = E_0(\hat{\mathbf{x}})$ and $\Delta E^{\text{sphere}}(\mathbf{x}) = \Delta E(\hat{\mathbf{x}})$ and the daily Fubini--Study step $\bar{\delta}_{\text{FS},t} = \arccos\lvert\langle\bar{\psi}_t\vert\bar{\psi}_{t-1}\rangle\rvert$ \cite{zanardi_2007}, where $\lvert\bar{\psi}_t\rangle$ is the principal eigenstate of the day-$t$ mean projector $\bar{\rho}_t = \lvert\mathcal{U}_t\rvert^{-1}\sum_{i\in\mathcal{U}_t}\rho_0(\mathbf{x}_{i,t})$.

Finally, we track the learned non-commutativity of operators through the normalized commutator ratio $\kappa_{\text{nc}}(\mathcal{A})$ and compare it with the analytical root-mean-square GUE random-matrix baseline (\ref{app:derivations}):
\begin{equation}\label{eq:kappa_nc}
\begin{aligned}
\kappa_{\text{nc}}(\mathcal{A}) &= \frac{\sum_{1 \le j < k \le p}\lVert[A_j, A_k]\rVert_F}{\sum_{1 \le j < k \le p}\lVert A_j\rVert_F\,\lVert A_k\rVert_F} \in \bigl[0,\,\sqrt{2}\bigr], \\
\kappa_{\text{nc}}^{\text{GUE}}(d) &= \sqrt{\frac{2}{d}\bigl(1 - d^{-2}\bigr)}.
\end{aligned}
\end{equation}
With $d = 8$, random GUE matrices give the theoretical value $\kappa_{\text{nc}}^{\text{GUE}}(8) = 0.4961$ (with finite-sample empirical mean $0.4933$), while commuting diagonal models satisfy $\kappa_{\text{nc}} \equiv 0$ identically in every run. The exact perturbation formulas, symmetry-reduced number of parameters and a Dirac-to-econometrics notation table \cite{li_2026} are given in \ref{app:derivations}. Synthetic non-commutative positive controls are given in the Supplementary Material (Section~S1).

\section{Data, Causal News Features, and Evaluation Protocol}\label{sec:data}

\subsection{MOEX Equity Panel and Lagged News Features}\label{subsec:data_panel}

Our sample covers shares traded on the Moscow Exchange (MOEX) from April 17, 2020 to April 17, 2025. After excluding 18 holiday records with zero closes, $T = 1{,}251$ active trading days remain. Prices have been retrospectively adjusted for all seven stock splits and reverse stock splits during this period, while maintaining the invariance of daily ruble turnover. On each trading day $t$, the investable universe $\mathcal{U}_t$ selects the \texttt{Top-50} most liquid shares based on the 63-day rolling median of ruble turnover among actively traded issues, with \texttt{Top-30} forming a tier of liquid blue chips.

The target forecast variable is the cross-section-centered five-day forward logarithmic return $y^{\text{cs}}_{i,t,5} = r_{i,t\to t+5} - \lvert\mathcal{U}_t\rvert^{-1}\sum_{j \in \mathcal{U}_t} r_{j,t\to t+5}$, where $r_{i,t\to t+5} = \ln(P^{\text{adj}}_{i,t+5} / P^{\text{adj}}_{i,t})$. From April 1, 2022 to March 27, 2025, each model is evaluated strictly on the same $37{,}048$ unfiltered out-of-sample stock-day observations across $T_{\text{OOS}} = 759$ trading sessions.

The market-only specification (\texttt{MKT}, $p_{\text{mkt}} = 12$) combines one-day, five-day, 21-day and 63-day stock returns, 21-day volatility, turnover shock, 252-day market beta, cross-sectional ranks of 21-day returns, volatility and average daily turnover, and 5-day IMOEX index return and 21-day index volatility. The market-plus-news specification (\texttt{M+N}, $p = 18$) adds six daily macroeconomic news indicators based on Lenta.ru summaries with time stamps strictly before the start of the trading day ($t_{\text{news}} \in [t_{\text{prev}}, t - 1]$): a 21-day news volume shock, polarity and uncertainty estimates from a dictionary compiled before 2020, and three truncated singular-value text factors retrained within each training window. This design without pre-trained language models eliminates chronological pre-training leakage \cite{lopez_lira_2025,khubiev_2025}. All variables are standardized by the moments of the expanding training window, and the limits of sensitivity to survivorship bias and ex-dividend dates are given in the Supplementary Material (Section~S3).

\begin{figure*}[t]
  \centering
  \begin{subfigure}[t]{0.485\textwidth}
    \includegraphics[width=\linewidth]{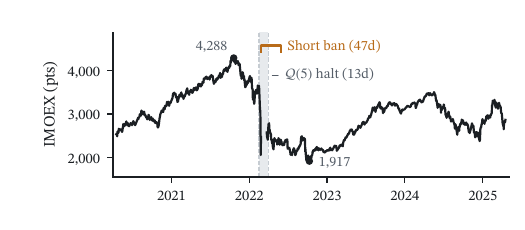}
    \caption{}
  \end{subfigure}\hfill
  \begin{subfigure}[t]{0.485\textwidth}
    \includegraphics[width=\linewidth]{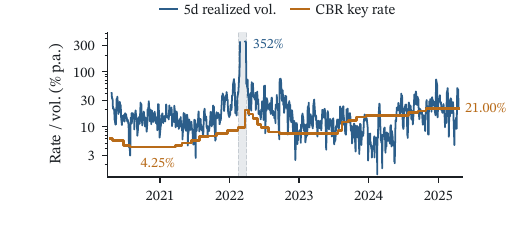}
    \caption{}
  \end{subfigure}

  \vspace{2.8pt}
  \begin{subfigure}[t]{0.99\textwidth}
    \includegraphics[width=\linewidth]{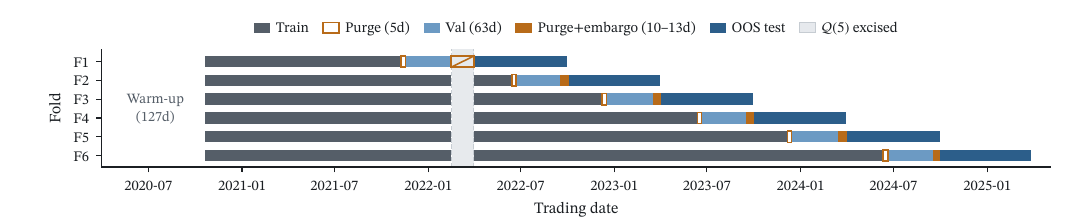}
    \caption{}
  \end{subfigure}

  \vspace{2.8pt}
  \begin{subfigure}[t]{0.485\textwidth}
    \includegraphics[width=\linewidth]{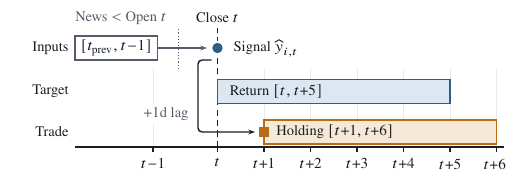}
    \caption{}
  \end{subfigure}\hfill
  \begin{subfigure}[t]{0.485\textwidth}
    \includegraphics[width=\linewidth]{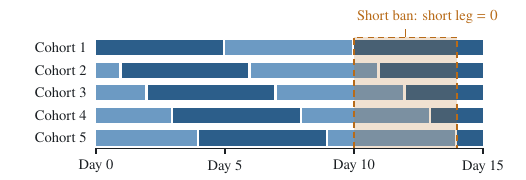}
    \caption{}
  \end{subfigure}
  \caption{MOEX market regimes, purged walk-forward calendar and causal execution protocol ($T = 1{,}251$ active days, 2020--2025). (\textbf{a})~IMOEX index (peak $4{,}288$, low $1{,}917$), isolated 13-day trading halt window in February--March 2022 $\mathcal{Q}(5)$ and 47-session CBR short selling ban. (\textbf{b})~Annualized 5-day volatility of the IMOEX index (peak $352\%$) and CBR key rate ($4.25\%$--$21.00\%$, OOS average $13.34\%$). (\textbf{c})~Six expanding windows \texttt{F1}--\texttt{F6} with 5-day training and validation purge, 63-day validation window and 10--13-day purge with pre-test embargo (open ocher frame on \texttt{F1} over excluded window $\mathcal{Q}(5)$). (\textbf{d})~Single forecast causal clock demarcates intersession news ($[t_{\text{prev}}, t - 1]$ before Open~$t$), Close-$t$ signal $\hat{y}_{i,t}$, 5-day target $[t, t + 5]$ and realized 1-day lag holding period $[t + 1, t + 6]$. (\textbf{e})~Five staggered weekly cohorts with a share of $1/5$ capital ($151$ rebalances per cohort) and a CBR short-ban mask for 38 rebalances.}
  \label{fig:moex_splits}
\end{figure*}

\subsection{Purged Walk-Forward Splits and Matched Model Ladder}\label{subsec:walkforward_ladder}

After a 127-day warm-up period, six semi-annual expanding windows (\texttt{F1}--\texttt{F6}; Fig.~\ref{fig:moex_splits}c) cover 125--129 out-of-sample sessions each, with a 5-day purge before the 63-day validation window and a 10--13-day purge with an embargo before the test. Such a buffer eliminates the overlap of five-day target returns. At the same time, 13 active sessions around the collapse in February 2022 and the closure of the exchange from February 28 to March 23 ($\mathcal{Q}(5)$) are not included in the training and validation \cite{kraevskiy_2024} and are allocated to the unclipped out-of-distribution crisis stress test ($320$ stock-day observations).

On identical sample splits and validation grids, we compare four matched model tiers \cite{shen_2026}:
\begin{itemize}\setlength\itemsep{1pt}\setlength\parskip{0pt}
  \item \textit{Tier 0 (parameter-free factors)}: expanding training average and standard cross-sectional signals \texttt{Mom5}, \texttt{LowVol}, \texttt{Mom21} and \texttt{Mom21-LowVol} \cite{jegadeesh_2001,ang_2006}.
  \item \textit{Tier 1 (classical learning models)}: \texttt{Ridge} ($13/19$ parameters on sets \texttt{MKT}/\texttt{M+N}), random Fourier regression \texttt{RFF-KRR} ($1{,}025$) \cite{rahimi_2007}, gradient boosting ensembles over trees \texttt{HistGBDT} and \texttt{LightGBM} ($960$ leaves) \cite{ke_2017,grinsztajn_2022}, neural networks \texttt{MLP-tanh} and \texttt{MLP-ReLU} ($827/1{,}221$) \cite{gu_2020}, and bounded-output model \texttt{FiLM-MLP (iso)} ($835/1{,}207$) \cite{perez_2018}.
  \item \textit{Tier 2 (variational quantum circuits)}: noiseless 3-qubit cyclic ($L = 3$, $155/200$ parameters) and affine ($L = 9$, $794/1{,}118$) VQCs with ring CNOT layers versus separable \texttt{NoCNOT} layers under full Hermitian and local Pauli-$Z/X$ readouts \cite{perez_salinas_2020,schuld_sweke_2021,cerezo_2021_bp,kobayashi_2023,schnabel_2025}.
  \item \textit{Tier 3 (matrix-form QCML and substitutions)}: \texttt{QCML Herm} ($d \in \{2,4,8,16\}$ and fold-selected $d^*_f$; $832/1{,}216$ parameters for $d = 8$), real-symmetric \texttt{QCML Real} ($d = 11$, $858/1{,}254$) and commuting \texttt{QCML Diag} ($d = 64$, $832/1{,}216$, selecting $\beta^*_f$ across $\{0.05,0.25,1,2,4,8,16\}$) \cite{samson_2024,candelori_2025}.
\end{itemize}
Deterministic models are trained once per window. Primary stochastic models average $N_{\text{seeds}} = 10$ random initializations ($N_{\text{seeds}} = 5$ in secondary ablations and $N_{\text{seeds}} = 3$ in daily spectral runs), capturing averages across single-seed runs and cross-seed ensembles. The out-of-sample score $R^2_{\text{OOS}}$ is evaluated strictly against the expanding training mean of each window, so that any constant prediction receives $R^2_{\text{OOS}} \equiv 0$ \cite{campbell_2008,gu_2020}.

\subsection{Statistical Inference and Realistic Portfolio Frictions}\label{subsec:inference_frictions}

The quality of forecasts is assessed by pooled daily Spearman rank correlation ($\overline{\mathrm{RankIC}}$), four-factor orthogonalized Fama--MacBeth $\mathrm{RankIC}_{\text{FM}}$ (orthogonalized against \texttt{Mom21}, \texttt{LowVol}, \texttt{Mom5} and turnover shock), worst-decile $\mathrm{Tail10}$ RMSE and $5\%$ pinball loss $\mathrm{PB}_{0.05}$. Pairwise model comparisons use Newey--West HAC standard errors ($L_{\text{HAC}} = 12$ lags; $p_{\text{NW}}$) \cite{newey_west_1987}, Harvey--Leybourne--Newbold Diebold--Mariano tests ($\mathrm{DM}$, $p_{\text{DM}}$) \cite{diebold_mariano_1995}, stationary block bootstrap ($2{,}000$ samples, average block $10$ days; $p_{\text{boot}}$), Holm--Bonferroni familywise correction ($p_{\text{Holm}}$) and Hansen's $95\%$ Model Confidence Set ($\mathrm{MCS}_{0.05}$; $p_{\text{MCS}}$) \cite{hansen_2011,li_2026}. When the 5-day index volatility exceeds the 90th percentile of the expanding training window, the sessions are causally classified as stress days. Crisis detection tests evaluate both contemporaneous and 5-day forward area under the receiver operating characteristic curve (AUROC), along with five-lag VAR(5) Granger causality tests ($W_{\text{Granger}}, p_{\text{Granger}}$) with and without five lags of $(\lVert\mathbf{x}\rVert_2, \lVert\mathbf{x}\rVert_2^2)$ \cite{hammond_2026}.

The weekly portfolio simulator has five overlapping $1/5$ capital cohorts ($151$ rebalances per cohort; Fig.~\ref{fig:moex_splits}d,e). Transactions based on close-$t$ signals are made at close $t + 1$, taking into account the drift of portfolio weights and the exact annualized one-way turnover ($\mathrm{TO}_{\text{ann}}$). Given one-way transaction costs $c_{\text{tc}} \in \{0, 5, 10, 15, 25, 50\}\text{ bps}$ (baseline $15\text{ bps}$ plus liquidity-differentiated scale $10/30\text{ bps}$), we track three indicators: (i) active long-only Information Ratio ($\mathrm{IR}_{\text{LO}}$) of the top quintile over the equal-weight \texttt{Top-50} universe; (ii) dollar-neutral long--short Sharpe ratio ($\mathrm{SR}_{\text{LS}}$) within \texttt{Top-30} with the short leg reset to zero during the 47-session CBR short selling ban ($38$ affected cohort rebalances) and with the $500\text{ bps/yr}$ borrowing rate outside the ban (with $0$ and $1{,}200\text{ bps/yr}$ in sensitivity tests); and (iii) stand-alone price-only long-only Sharpe above the official CBR key rate (average $13.34\%$). Selection bias for all strategies is controlled through the Deflated Sharpe Ratio ($\mathrm{DSR}$, threshold $0.95$) with $K_{\text{trials}} = 40$ effective trials on $70$ estimated portfolio series and autocorrelation-corrected effective sample length $T_{\text{eff}}$ \cite{bailey_2014,harvey_2016}.

\section{Empirical Results}\label{sec:results}

\subsection{Predictive Benchmark and Empirical De-Quantization}\label{subsec:dequant}

\begin{figure*}[t]
  \centering
  \begin{subfigure}[t]{0.485\textwidth}
    \includegraphics[width=\linewidth]{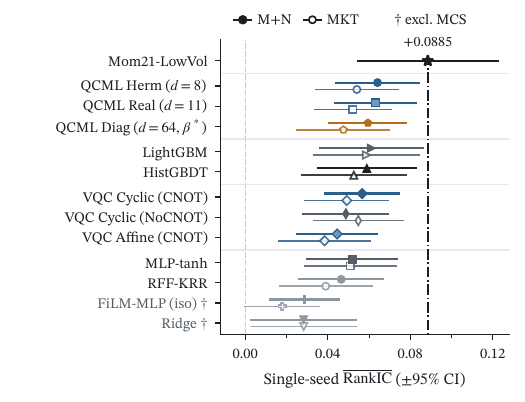}
    \caption{}
  \end{subfigure}\hfill
  \begin{subfigure}[t]{0.485\textwidth}
    \includegraphics[width=\linewidth]{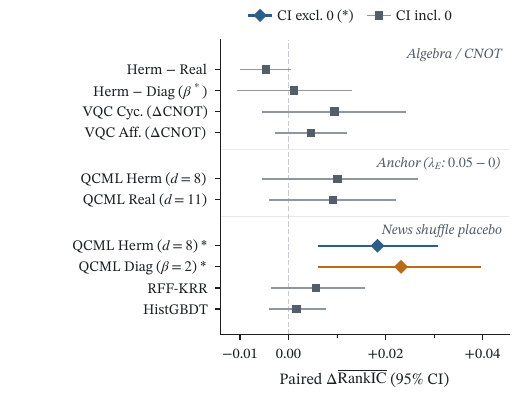}
    \caption{}
  \end{subfigure}

  \vspace{2.8pt}
  \begin{subfigure}[t]{0.485\textwidth}
    \includegraphics[width=\linewidth]{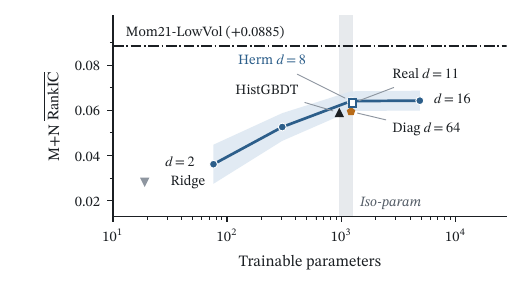}
    \caption{}
  \end{subfigure}\hfill
  \begin{subfigure}[t]{0.485\textwidth}
    \includegraphics[width=\linewidth]{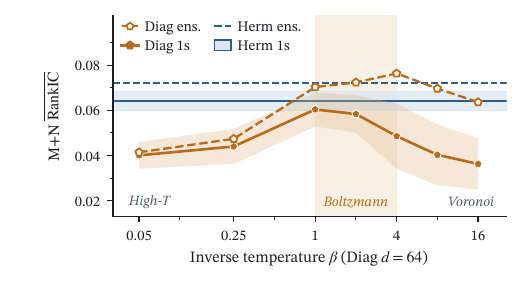}
    \caption{}
  \end{subfigure}
  \caption{Cross-sectional out-of-sample benchmark, pairwise comparisons and statistical-mechanical scaling ($37{,}048$ rows, $759$ days; $N_{\text{seeds}}=10$ main models, $5$ auxiliary ablations). (\textbf{a})~Values of single-seed $\overline{\mathrm{RankIC}}$ ($\pm 95\%$ NW-HAC CI) for all $13$ models in Table~\ref{tab:headline_oos} on \texttt{M+N} (filled) and \texttt{MKT} (open) relative to the zero-parameter ceiling \texttt{Mom21-LowVol} ($+0.0885$); daggers ($\dagger$) mark the two models excluded from Hansen's $95\%$ $\mathrm{MCS}_{0.05}^{\mathrm{IC}}$. (\textbf{b})~Stationary block-bootstrap $95\%$ CIs for pairwise \texttt{M+N} contrasts grouped by hypothesis (diamonds mark intervals excluding zero). (\textbf{c})~Single-seed \texttt{M+N} Hilbert-dimension frontier ($d \in \{2,4,8,16\}$, $\pm 1\sigma_{\text{seed}}$) and iso-parameter band ($960$--$1{,}254$ parameters) with saturation at $d=8$. (\textbf{d})~Seven-temperature sweep for commuting Gibbs--RBF (\texttt{QCML Diag}, $d=64$, $\beta \in [0.05, 16]$; ensemble and single-seed $\pm 1\sigma_{\text{seed}}$) versus \texttt{QCML Herm} ($d=8$), from high-temperature underfitting through smooth Boltzmann parity ($\beta \in [1, 4]$) to Voronoi freezing at $\beta = 16$.}
  \label{fig:dequant}
\end{figure*}

Table~\ref{tab:headline_oos} and Fig.~\ref{fig:dequant} summarize out-of-sample forecasting of five-day returns for all $37{,}048$ stock-days ($759$ trading sessions). Paired statistical tests are reported in Table~\ref{tab:dequant_contrasts}. Four results stand out.

\begin{table*}[t]
\centering
\caption{Out-of-sample cross-sectional forecasting benchmark ($h=5$, $37{,}048$ unfiltered stock--day rows, $759$ trading days, one-day execution lag, $15\text{ bps}$ one-way transaction costs). Stochastic models undergo $N_{\text{seeds}}=10$ independent initializations; daily Spearman rank correlations ($\overline{\mathrm{RankIC}}$) for single-seed average (\texttt{1s}) and ensemble of ten runs (\texttt{Ens}) are given for market-only (\texttt{MKT}, $p_{\text{mkt}}=12$) and market-news (\texttt{M+N}, $p=18$) inputs. Economic columns give ensemble Campbell--Thompson $R^2_{\text{OOS}}$ on \texttt{M+N} against expanding fold training averages, top-quintile active long-only information ratio over equal-weight equity universe ($\mathrm{IR}_{\text{LO}}$), break-even one-way costs ($c^*$) and Deflated Sharpe Ratio ($\mathrm{DSR}_{\text{LO}}$ at $K_{\text{trials}}=40$). Single-seed economic indicators and long--short Sharpe ratios are presented in Table~\ref{tab:economics_ood}. Rank-only \texttt{Mom21-LowVol} scores are not scaled ($R^2_{\text{OOS}}$ is marked n/a).}
\label{tab:headline_oos}
\footnotesize
\setlength{\tabcolsep}{4.5pt}
\resizebox{\textwidth}{!}{%
\begin{tabular}{@{}lccccccc@{}}
\toprule
Model & Params (\texttt{MKT}/\texttt{M+N}) & \texttt{MKT} \texttt{1s}/\texttt{Ens} $\overline{\mathrm{RankIC}}$ & \texttt{M+N} \texttt{1s}/\texttt{Ens} $\overline{\mathrm{RankIC}}$ & \texttt{M+N} $R^2_{\text{OOS}}$ & $\mathrm{IR}_{\text{LO}}$ ($15\text{ bps}$) & $c^*$ (bps) & $\mathrm{DSR}_{\text{LO}}$ \\
\midrule
\texttt{Mom21-LowVol} & $0 / 0$ & $+0.0885 / +0.0885$ & $+0.0885 / +0.0885$ & n/a & $+0.488$ & $30.4$ & $0.087$ \\
\texttt{Ridge} & $13 / 19$ & $+0.0283 / +0.0283$ & $+0.0283 / +0.0283$ & $+0.00114$ & $+0.133$ & $18.4$ & $0.026$ \\
\texttt{RFF-KRR} & $1{,}025 / 1{,}025$ & $+0.0390 / +0.0418$ & $+0.0465 / +0.0524$ & $+0.00415$ & $+0.754$ & $33.0$ & $0.201$ \\
\texttt{HistGBDT} & $960 / 960$ & $+0.0526 / +0.0526$ & $+0.0589 / +0.0589$ & $+0.00274$ & $+1.072$ & $39.2$ & $0.519$ \\
\texttt{LightGBM} & $960 / 960$ & $+0.0588 / +0.0588$ & $+0.0611 / +0.0611$ & $+0.00500$ & $+0.969$ & $36.0$ & $0.320$ \\
\texttt{MLP-tanh} & $827 / 1{,}221$ & $+0.0509 / +0.0574$ & $+0.0518 / +0.0602$ & $+0.00548$ & $+0.741$ & $31.1$ & $0.210$ \\
\texttt{FiLM-MLP (iso)} & $835 / 1{,}207$ & $+0.0179 / +0.0195$ & $+0.0287 / +0.0324$ & $-0.00282$ & $+0.684$ & $27.7$ & $0.223$ \\
\midrule
\texttt{VQC Cyclic (CNOT)} & $155 / 200$ & $+0.0492 / +0.0568$ & $+0.0567 / +0.0656$ & $+0.00512$ & $+1.214$ & $40.8$ & $0.637$ \\
\texttt{VQC Cyclic (NoCNOT)} & $155 / 200$ & $+0.0548 / +0.0604$ & $+0.0488 / +0.0561$ & $+0.00465$ & $+1.067$ & $37.8$ & $0.485$ \\
\texttt{VQC Affine (CNOT)} & $794 / 1{,}118$ & $+0.0384 / +0.0433$ & $+0.0445 / +0.0526$ & $+0.00206$ & $+1.178$ & $40.0$ & $0.686$ \\
\midrule
\textbf{\texttt{QCML Herm} ($d=8$)} & $832 / 1{,}216$ & $+0.0541 / +0.0610$ & $+0.0641 / +0.0723$ & $+0.00519$ & $+1.026$ & $37.7$ & $0.415$ \\
\textbf{\texttt{QCML Real} ($d=11$)} & $858 / 1{,}254$ & $+0.0521 / +0.0595$ & $+0.0632 / +0.0768$ & $+0.00543$ & $+1.114$ & $39.0$ & $0.578$ \\
\textbf{\texttt{QCML Diag} ($d=64$, $\beta^*$)} & $832 / 1{,}216$ & $+0.0476 / +0.0512$ & $+0.0595 / +0.0711$ & $+0.00604$ & $+0.845$ & $34.1$ & $0.311$ \\
\bottomrule
\end{tabular}%
}
\end{table*}

First, the zero-parameter momentum--volatility factor \texttt{Mom21-LowVol} sets the ranking ceiling: $\overline{\mathrm{RankIC}} = +0.0885$ ($t_{\text{NW}} = +5.04$). Among the trained models on \texttt{M+N}, \texttt{QCML Herm} ($d = 8$) achieves the highest single-seed average ($+0.0641 \pm 0.0043$) and rises to $+0.0723$ ($t_{\text{NW}} = +6.30$) in an ensemble of ten runs, but Hansen's $95\%$ Model Confidence Set ($\mathrm{MCS}_{0.05}^{\mathrm{IC}}$) retains all 11 remaining Table~\ref{tab:headline_oos} models (grouping \texttt{QCML Herm}, \texttt{QCML Real}, \texttt{QCML Diag}, \texttt{LightGBM}, \texttt{HistGBDT}, \texttt{MLP-tanh} and \texttt{VQC Cyclic (CNOT)} at $p_{\text{MCS}} = 0.653$) and excludes only \texttt{Ridge} and \texttt{FiLM-MLP (iso)} ($p_{\text{MCS}} = 0.001$). Direct comparison with \texttt{HistGBDT} confirms statistical parity in rank correlation ($\Delta = +0.0133$, $p_{\text{NW}} = 0.3121$; HLN-$\mathrm{DM} = +1.01$, $p_{\text{DM}} = 0.3150$), squared error ($\mathrm{DM} = +0.71$, $p_{\text{DM}} = 0.4766$) and tail loss ($\mathrm{Tail10} = 0.1145$ vs.\ $0.1141$; $\mathrm{PB}_{0.05} = 0.0147$ vs.\ $0.0149$). As the Hilbert dimension $d \in \{2,4,8,16\}$ increases, the ranking rises from $d = 2$ and saturates at $d = 8$ and $d = 16$ in ensembles ($+0.0456 \to +0.0723$ and $+0.0726$) and single-seed runs ($+0.0362 \to +0.0641$ and $+0.0643$; Fig.~\ref{fig:dequant}c). Per-fold $d^*_f$ tuning doubles inter-seed dispersion without improving single-seed accuracy ($+0.0645 \pm 0.0091$). After orthogonalization against four standard risk factors ($\mathrm{RankIC}_{\text{FM}}$), boosted trees (\texttt{LightGBM} $+0.0434$, \texttt{HistGBDT} $+0.0394$) and \texttt{QCML Real} ($+0.0353$) outperform \texttt{QCML Herm} ($+0.0305$). Deterministic \texttt{LightGBM} leads single-seed $R^2_{\text{OOS}}$ ($+0.00500$ vs.\ \texttt{QCML Herm} $+0.00399$), while averaging over ten runs puts \texttt{QCML Diag} ($+0.00604$) and \texttt{MLP-tanh} ($+0.00548$) above \texttt{QCML Herm} ($+0.00519$; Table~\ref{tab:headline_oos}).

\begin{table*}[t]
\centering
\caption{Paired daily $\overline{\mathrm{RankIC}}$ contrasts ($759$ trading days) test empirical de-quantization, VQC entanglement, and conditioning on lagged news. Two-sided $p$-values are reported for Newey--West HAC ($L_{\text{HAC}}=12$, $p_{\text{NW}}$) and stationary block bootstrap ($N_{\text{boot}}=2{,}000$, $\bar{b}=10$, $p_{\text{boot}}$). For Panel~A and QCML entries in Panel~C, Holm--Bonferroni $p_{\text{Holm}}$ corrections are given for the entire family of twelve \texttt{M+N} contrasts (including the difference \texttt{QCML Real} $\lambda_E=0.05$ minus $\lambda_E=0$: $\Delta=+0.0092$, $p_{\text{NW}}=0.1782$, $p_{\text{Holm}}=1.0000$).}
\label{tab:dequant_contrasts}
\footnotesize
\renewcommand{\arraystretch}{1.12}
\setlength{\tabcolsep}{5pt}
\resizebox{\textwidth}{!}{%
\begin{tabular}{@{}lccccc@{}}
\toprule
Contrast (first minus second) & Auxiliary diagnostic & $\Delta\overline{\mathrm{RankIC}}$ & $p_{\text{NW}}$ & $p_{\text{boot}}$ & $p_{\text{Holm}}$ \\
\midrule
\multicolumn{6}{@{}l}{\textit{Panel A: Operator-swap and baseline contrasts (\texttt{M+N}, $10$-seed ensembles, \texttt{QCML Herm} $d=8$ reference)}} \\
\texttt{QCML Herm} ($d=8$) $-$ \texttt{QCML Real} ($d=11$, $F_{jk}\equiv 0$) & $\kappa_{\text{nc}} = 0.2999$ vs.\ $0.2519$ & $-0.0046$ & $0.0917$ & $0.0885$ & $0.6420$ \\
\texttt{QCML Herm} ($d=8$) $-$ \texttt{QCML Diag} ($d=64$, $\beta^*$, $\kappa_{\text{nc}}\equiv 0$) & $\sigma_{\text{seed}} = 0.0043$ vs.\ $0.0083$ & $+0.0011$ & $0.8532$ & $0.8480$ & $1.0000$ \\
\texttt{QCML Herm} ($\lambda_E=0.05$) $-$ \texttt{QCML Herm} ($\lambda_E=0$) & $\kappa_{\text{nc}} = 0.2999$ vs.\ $0.4818$ & $+0.0101$ & $0.2069$ & $0.2110$ & $1.0000$ \\
\texttt{QCML Herm} ($d=8$) $-$ \texttt{HistGBDT} & $\mathrm{MCS}_{0.05}^{\mathrm{IC}}$ tie & $+0.0133$ & $0.3121$ & $0.2835$ & $1.0000$ \\
\texttt{QCML Herm} ($d=8$) $-$ \texttt{MLP-tanh} & $\mathrm{MCS}_{0.05}^{\mathrm{IC}}$ tie & $+0.0121$ & $0.3300$ & $0.3025$ & $1.0000$ \\
\texttt{QCML Herm} ($d=8$) $-$ \texttt{RFF-KRR} & Static RBF control & $+0.0199$ & $0.0377$ & $0.0365$ & $0.3264$ \\
\texttt{QCML Herm} ($d=8$) $-$ \texttt{VQC Affine (CNOT)} & Iso-param VQC control & $+0.0197$ & $0.0363$ & $0.0300$ & $0.3264$ \\
\texttt{QCML Herm} ($d=8$) $-$ \texttt{FiLM-MLP (iso)} & Bounded classical control & $+0.0399$ & $0.0001$ & $<0.0005$ & $0.0009$ \\
\texttt{QCML Herm} ($d=8$) $-$ \texttt{Mom21-LowVol} & Factor ceiling & $-0.0162$ & $0.2957$ & $0.3040$ & $1.0000$ \\
\midrule
\multicolumn{6}{@{}l}{\textit{Panel B: Three-qubit VQC entanglement contrasts (\texttt{CNOT} minus \texttt{NoCNOT}; $Q_{\mathrm{MW}}$ vs.\ $\le 10^{-6}$)}} \\
\texttt{MKT} Affine, full-Hermitian ($N_{\text{seeds}}=10$) & $Q_{\text{MW}} = 0.378$ & $-0.0007$ & $0.8133$ & $0.8105$ & --- \\
\texttt{MKT} Affine, local-$Z/X$ ($N_{\text{seeds}}=5$) & $Q_{\text{MW}} = 0.389$ & $+0.0003$ & $0.9490$ & $0.9450$ & --- \\
\texttt{MKT} Cyclic, full-Hermitian ($N_{\text{seeds}}=10$) & $Q_{\text{MW}} = 0.468$ & $-0.0036$ & $0.6008$ & $0.5950$ & --- \\
\texttt{M+N} Affine, full-Hermitian ($N_{\text{seeds}}=10$) & $Q_{\text{MW}} = 0.425$ & $+0.0047$ & $0.1881$ & $0.2085$ & --- \\
\texttt{M+N} Affine, local-$Z/X$ ($N_{\text{seeds}}=5$) & $Q_{\text{MW}} = 0.433$ & $+0.0049$ & $0.4508$ & $0.4285$ & --- \\
\texttt{M+N} Cyclic, full-Hermitian ($N_{\text{seeds}}=10$) & $Q_{\text{MW}} = 0.532$ & $+0.0095$ & $0.2279$ & $0.2210$ & --- \\
\midrule
\multicolumn{6}{@{}l}{\textit{Panel C: Date-wise news-shuffle placebos (real vs.\ 5-seed shuffled news, \texttt{M+N}; deterministic \texttt{HistGBDT}, 10-seed real stochastic ensembles)}} \\
\texttt{QCML Herm} ($d=8$) & Projective ground state & $+0.0184$ & $0.0071$ & $0.0045$ & $0.0776$ \\
\texttt{QCML Diag} ($d=64$, $\beta=2$) & Commuting Gibbs kernel & $+0.0232$ & $0.0080$ & $0.0060$ & $0.0797$ \\
\texttt{HistGBDT} & Boosted decision trees & $+0.0016$ & $0.5838$ & $0.5740$ & --- \\
\texttt{RFF-KRR} & Static random Fourier & $+0.0057$ & $0.2741$ & $0.2590$ & --- \\
\bottomrule
\end{tabular}%
}
\end{table*}

Second, matched operator substitutions under $1/d$ regularization de-quantize the complex phase and noncommutativity of the matrices (Table~\ref{tab:dequant_contrasts}A, Fig.~\ref{fig:dequant}b). The real-symmetric \texttt{QCML Real} ($d = 11$, zero Berry curvature) is slightly ahead of \texttt{QCML Herm} in ensembles of ten runs ($+0.0768$ vs.\ $+0.0723$, $p_{\text{NW}} = 0.0917$; single-seed $+0.0632$ vs.\ $+0.0641$). The classical commuting Gibbs--RBF swap \texttt{QCML Diag} ($d = 64$, $\beta^*_f \in \{1, 4\}$) is comparable to \texttt{QCML Herm} in ensemble ranking ($+0.0711$, $\Delta = +0.0011$, $p_{\text{NW}} = 0.8532$), but has almost twice the inter-run spread ($\sigma_{\text{seed}} = 0.0083$ vs.\ $0.0043$). In a seven-temperature sweep ($\beta \in [0.05, 16]$; Fig.~\ref{fig:dequant}d), \texttt{QCML Diag} matches \texttt{QCML Herm} in the smooth Boltzmann regime $\beta \in [1, 4]$ (ensemble $+0.0724$ at $\beta = 2$), before freezing to a Voronoi plateau at $\beta = 16$, where the single-seed $\overline{\mathrm{RankIC}}$ drops from $+0.0604$ at $\beta = 1$ to $+0.0363$. Energy anchoring ($\lambda_E = 0.05$) lowers the trained non-commutativity $\kappa_{\text{nc}} = 0.2999$ by $39\%$ below the random level of GUE ($0.4933$). Without an anchor ($\lambda_E = 0$), the $\kappa_{\text{nc}}$ value remains near the random level ($0.4818$). Anchoring does not increase the accuracy of the ensemble ($p_{\text{NW}} = 0.2069$), but works as a regularizer, reducing the spread of runs of the Hermitian model by more than half ($\sigma_{\text{seed}} = 0.0043$ vs.\ $0.0094$).

Third, three-qubit VQCs do not benefit from quantum entanglement (Table~\ref{tab:dequant_contrasts}B, Fig.~\ref{fig:dequant}b) \cite{schnabel_2025,bowles_2024}. In all six matched comparisons of \texttt{CNOT} vs. \texttt{NoCNOT}, the Meyer--Wallach entanglement measure \cite{meyer_wallach_2002} increases from $\le 10^{-6}$ to $0.378$--$0.532$. However, none of the six accuracy differences are statistically significant (minimum $p_{\text{NW}} = 0.1881$).

Fourth, placebo tests with news shuffling by date support the role of lagged news (Table~\ref{tab:dequant_contrasts}C, Fig.~\ref{fig:dequant}b). Lagged news helps learned prototype and subspace gating without requiring non-commuting operators: real minus date-shuffled news brings $\Delta\overline{\mathrm{RankIC}} = +0.0184$ in \texttt{QCML Herm} ($p_{\text{NW}} = 0.0071$, $p_{\text{Holm}} = 0.0776$) and $+0.0232$ in commuting \texttt{QCML Diag} ($p_{\text{NW}} = 0.0080$, $p_{\text{Holm}} = 0.0797$). Static random-feature \texttt{RFF-KRR} ($+0.0057$, $p_{\text{NW}} = 0.2741$) and shallow \texttt{HistGBDT} ($+0.0016$, $p_{\text{NW}} = 0.5838$) do not show significant sensitivity to placebo.

\subsection{Refutation of Spectral-Gap Crisis Detection}\label{subsec:spectral}

\begin{figure*}[t]
  \centering
  \begin{subfigure}[t]{0.265\textwidth}
    \includegraphics[width=\linewidth]{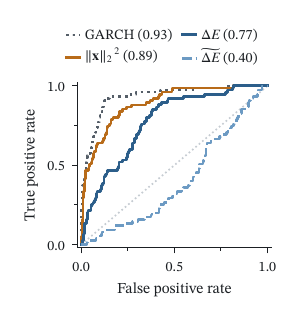}
    \caption{}
  \end{subfigure}\hfill
  \begin{subfigure}[t]{0.305\textwidth}
    \includegraphics[width=\linewidth]{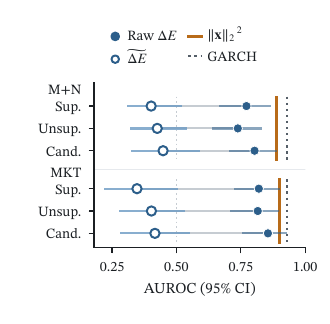}
    \caption{}
  \end{subfigure}\hfill
  \begin{subfigure}[t]{0.200\textwidth}
    \includegraphics[width=\linewidth]{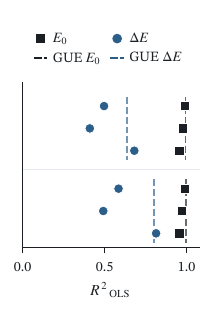}
    \caption{}
  \end{subfigure}\hfill
  \begin{subfigure}[t]{0.200\textwidth}
    \includegraphics[width=\linewidth]{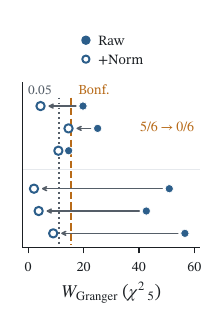}
    \caption{}
  \end{subfigure}
  \caption{Spectral-gap norm-confound refutation on $759$ out-of-sample days ($N_{\text{seeds}} = 3$ per regime). (\textbf{a})~Contemporaneous stress ROC curves on supervised \texttt{M+N}: \texttt{GARCH(1,1)} ($0.930$) and $\lVert\mathbf{x}\rVert_2^2$ ($0.888$) dominate the raw $\Delta E$ ($0.772$), while the scale-free $\widetilde{\Delta E}$ falls below chance ($0.402$). (\textbf{b})~Crisis AUROC ($95\%$ bootstrap CI) drops from raw $\Delta E$ ($0.738$--$0.855$) to scale-free $\widetilde{\Delta E}$ ($0.347$--$0.448$) in all six modality--training cells. (\textbf{c})~Daily $R^2_{\text{OLS}}$ regressions of $E_0$ on $\lVert\mathbf{x}\rVert_2^2$ and $\Delta E$ on $[1, \lVert\mathbf{x}\rVert_2, \lVert\mathbf{x}\rVert_2^2]$ along with ten untrained GUE($8$) random-matrix baselines. (\textbf{d})~Forward VAR(5) Granger Wald statistics $W_{\text{Granger}}$ drop from $5/6$ to $0/6$ at the Bonferroni threshold $\alpha/6 = 0.0083$ ($\chi^2_5 = 15.53$; from $6/6$ to $1/6$ at nominal $\alpha = 0.05$) after controlling for five lags of $(\lVert\mathbf{x}\rVert_2, \lVert\mathbf{x}\rVert_2^2)$.}
  \label{fig:spectral}
\end{figure*}

Table~\ref{tab:spectral_audit} and Fig.~\ref{fig:spectral} check whether QCML spectral gaps reveal high-volatility market stress on $759$ out-of-sample days ($684$ normal, $75$ stress). We compare the supervised ($\lambda_E = 0.05$), unsupervised ($\lambda_E = 1.0$) and variance-weighted Candelori ($w_{\text{var}} = 0.1$) models with ten random untrained GUE($d = 8$) samples, the squared variable norm $\lVert\mathbf{x}\rVert_2^2$ and a standard \texttt{GARCH(1,1)} filter \cite{candelori_2025,hammond_2026}.

\begin{table*}[t]
\centering
\caption{Audit of spectral crisis detection with norm control over $759$ out-of-sample trading days ($684$ normal, $75$ stress; $N_{\text{seeds}}=3$ per QCML training regime; $10$ untrained GUE($d=8$) draws). Daily linear $R^2_{\text{OLS}}(E_0)$ regresses $E_0$ on $\lVert\mathbf{x}\rVert_2^2$, and quadratic $R^2_{\text{OLS}}(\Delta E)$ regresses $\Delta E$ on $[1, \lVert\mathbf{x}\rVert_2, \lVert\mathbf{x}\rVert_2^2]$; columns~4--9 list contemporaneous crisis AUROC for raw $\Delta E$, scale-free $\widetilde{\Delta E}$, unit-sphere $\Delta E^{\text{sphere}}$ and $E_0^{\text{sphere}}$, Fubini--Study step $\bar{\delta}_{\text{FS},t}$ and fold-causal quadratic residual $\Delta E^{\perp}$. Forward VAR(5) Granger Wald statistics ($W_{\text{Granger}}$) and $p$-values ($729$ lag-aligned days) compare raw $\Delta E$ with norm-controlled regressions adding five lags of both $\lVert\mathbf{x}\rVert_2$ and $\lVert\mathbf{x}\rVert_2^2$.}
\label{tab:spectral_audit}
\footnotesize
\setlength{\tabcolsep}{3.5pt}
\resizebox{\textwidth}{!}{%
\begin{tabular}{@{}lcccccccccc@{}}
\toprule
Specification & $R^2_{\text{OLS}}(E_0)$ & $R^2_{\text{OLS}}(\Delta E)$ & $\Delta E$ & $\widetilde{\Delta E}$ & $\Delta E^{\text{sphere}}$ & $E_0^{\text{sphere}}$ & $\bar{\delta}_{\text{FS},t}$ & $\Delta E^{\perp}$ & Raw $W_{\text{Granger}}$ ($p_{\text{Granger}}$) & Ctrl $W_{\text{Granger}}$ ($p_{\text{Granger}}$) \\
\midrule
\multicolumn{11}{@{}l}{\textit{Baselines and untrained random-matrix null ($759$ days)}} \\
\texttt{GARCH(1,1)} (\texttt{M+N}/\texttt{MKT}) & --- & --- & $0.930$ & --- & --- & --- & --- & --- & --- & --- \\
$\lVert\mathbf{x}\rVert_2^2$ (\texttt{M+N}/\texttt{MKT}) & --- & --- & $0.888 / 0.900$ & --- & --- & --- & --- & --- & --- & --- \\
Untrained \texttt{GUE} ($d=8$, \texttt{M+N}) & $0.9929$ & $0.6359$ & --- & --- & --- & --- & --- & --- & --- & --- \\
Untrained \texttt{GUE} ($d=8$, \texttt{MKT}) & $0.9953$ & $0.8015$ & --- & --- & --- & --- & --- & --- & --- & --- \\
\midrule
\multicolumn{11}{@{}l}{\textit{Trained matrix-QCML ($d=8$, $N_{\text{seeds}}=3$) on market-plus-news (\texttt{M+N})}} \\
Supervised ($\lambda_E = 0.05$) & $0.9911$ & $0.4962$ & $0.772$ & $0.402$ & $0.514$ & $0.756$ & $0.431$ & $0.183$ & $19.79$ ($0.0014$) & $4.45$ ($0.4863$) \\
Unsupervised ($\lambda_E = 1.0$) & $0.9769$ & $0.4094$ & $0.738$ & $0.425$ & $0.310$ & $0.737$ & $0.497$ & $0.157$ & $25.09$ ($0.0001$) & $14.57$ ($0.0124$) \\
Candelori ($w_{\text{var}} = 0.1$) & $0.9551$ & $0.6806$ & $0.803$ & $0.448$ & $0.447$ & $0.860$ & $0.401$ & $0.145$ & $14.61$ ($0.0121$) & $10.90$ ($0.0535$) \\
\midrule
\multicolumn{11}{@{}l}{\textit{Trained matrix-QCML ($d=8$, $N_{\text{seeds}}=3$) on market-only (\texttt{MKT})}} \\
Supervised ($\lambda_E = 0.05$) & $0.9892$ & $0.5846$ & $0.820$ & $0.347$ & $0.248$ & $0.727$ & $0.490$ & $0.148$ & $50.98$ ($<10^{-4}$) & $2.08$ ($0.8374$) \\
Unsupervised ($\lambda_E = 1.0$) & $0.9722$ & $0.4914$ & $0.816$ & $0.403$ & $0.237$ & $0.901$ & $0.503$ & $0.139$ & $42.67$ ($<10^{-4}$) & $3.77$ ($0.5824$) \\
Candelori ($w_{\text{var}} = 0.1$) & $0.9552$ & $0.8135$ & $0.855$ & $0.416$ & $0.660$ & $0.911$ & $0.599$ & $0.150$ & $56.62$ ($<10^{-4}$) & $9.03$ ($0.1078$) \\
\bottomrule
\end{tabular}%
}
\end{table*}

The random-matrix null model immediately confirms the Weyl scaling mechanism from Section~\ref{subsec:norm_scaling}. Without any training on market data, untrained random GUE($d = 8$) matrices achieve $R^2_{\text{OLS}} > 0.99$ between $E_0$ and $\lVert\mathbf{x}\rVert_2^2$, and a simple quadratic dependence on $\lVert\mathbf{x}\rVert_2$ explains $64\%$ (\texttt{M+N}) to $80\%$ (\texttt{MKT}) of the daily variation in $\Delta E$ for untrained matrices (Fig.~\ref{fig:spectral}c). Since raw $\Delta E$ mainly tracks the magnitude of input features, its crisis AUROC ($0.738$--$0.803$ on \texttt{M+N}; $0.816$--$0.855$ on \texttt{MKT}) trails both plain $\lVert\mathbf{x}\rVert_2^2$ ($0.888$--$0.900$) and \texttt{GARCH(1,1)} ($0.930$; Fig.~\ref{fig:spectral}a,b).

When the radial scale is eliminated via Eq.~\eqref{eq:gap_scalefree}, the crisis separation disappears completely: the scale-free $\widetilde{\Delta E}$ falls below chance in all six trained configurations ($0.347$--$0.448$; Fig.~\ref{fig:spectral}a,b), and its daily correlation with $\lVert\mathbf{x}\rVert_2$ on supervised \texttt{M+N} drops from $r_{\text{level}} = +0.70$ to $-0.03$ (Fig.~\ref{fig:algo_geom}d,e). The same collapse continues five days ahead ($S^{\text{lead}}_{t\to t+5}$): $\widetilde{\Delta E}$ falls to $0.370$--$0.496$, while $\lVert\mathbf{x}\rVert_2^2$ ($0.820$) and \texttt{GARCH(1,1)} ($0.767$) maintain strong separation. Unit-sphere spectral gaps $\Delta E^{\text{sphere}}$ and Fubini--Study steps $\bar{\delta}_{\text{FS},t}$ remain near or below chance in five of the six configurations, peaking at $0.660$ and $0.599$ for \texttt{MKT} Candelori (still well below $\lVert\mathbf{x}\rVert_2^2 = 0.900$; Table~\ref{tab:spectral_audit}), and the $\Delta E^{\perp}$ regression residuals invert to $0.139$--$0.183$, since pre-crisis quadratic polynomials in $\lVert\mathbf{x}\rVert_2$ over-extrapolate at extreme crisis norms. In contrast, the unit-sphere ground energy $E_0^{\text{sphere}}$ retains directional sensitivity to stress across all objectives ($0.727$--$0.911$ contemporaneously), strongest for Candelori training ($0.860$--$0.911$ contemporaneous; $0.846$--$0.852$ five days ahead) and for market-only unsupervised training ($0.901$; $0.848$ five days ahead), remaining below the contemporaneous \texttt{GARCH(1,1)} ($0.930$) but exceeding its five-day-ahead score ($0.767$). Forward VAR(5) Granger tests on $729$ lag-aligned transitions confirm this conclusion (Fig.~\ref{fig:spectral}d) \cite{hammond_2026}: raw links $\Delta E \to \text{volatility}$ are significant in $5/6$ configurations at the Bonferroni threshold $\alpha/6 = 0.0083$ ($\chi^2_5 = 15.53$; $6/6$ at nominal $\alpha = 0.05$, $p_{\text{Granger}} \le 0.0121$), yet drop to $0/6$ Bonferroni-significant ($1/6$ nominal, $p_{\text{Granger}} = 0.0124$) after adding five lags of $(\lVert\mathbf{x}\rVert_2, \lVert\mathbf{x}\rVert_2^2)$ as controls.

\subsection{Portfolio Frictions and Crisis Forecast Clamping}\label{subsec:economics}

\begin{figure*}[t]
  \centering
  \begin{subfigure}[t]{0.485\textwidth}
    \includegraphics[width=\linewidth]{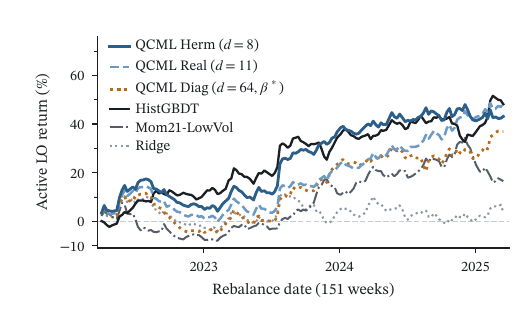}
    \caption{}
  \end{subfigure}\hfill
  \begin{subfigure}[t]{0.485\textwidth}
    \includegraphics[width=\linewidth]{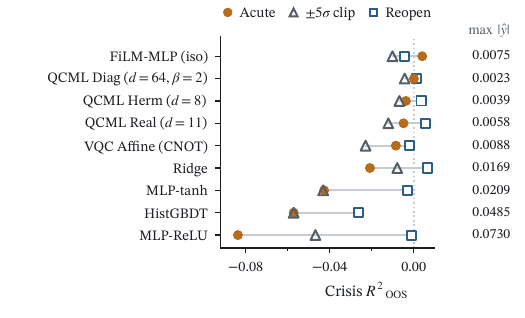}
    \caption{}
  \end{subfigure}

  \vspace{2.8pt}
  \begin{subfigure}[t]{0.420\textwidth}
    \includegraphics[width=\linewidth]{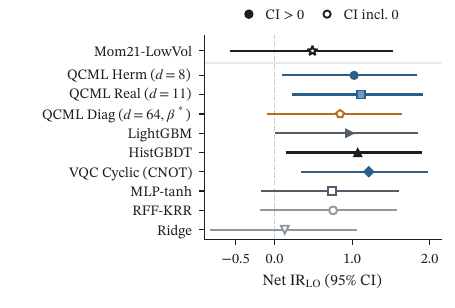}
    \caption{}
  \end{subfigure}\hfill
  \begin{subfigure}[t]{0.275\textwidth}
    \includegraphics[width=\linewidth]{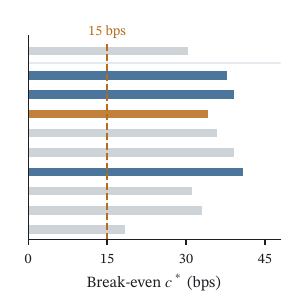}
    \caption{}
  \end{subfigure}\hfill
  \begin{subfigure}[t]{0.275\textwidth}
    \includegraphics[width=\linewidth]{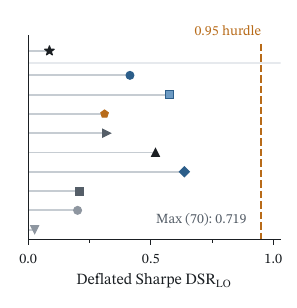}
    \caption{}
  \end{subfigure}
  \caption{Causal portfolio frictions on MOEX and February--March 2022 crisis shock bounds (\texttt{M+N}, one-day execution lag, $38$ masked short-ban rebalances, $500\text{ bps/yr}$ borrowing, $N_{\text{seeds}}=10$ main models). (\textbf{a})~Compounded net active long-only returns over an equal-weighted equity universe at $15\text{ bps}$ one-way costs for six representative trajectories ($151$ weekly cycles across five cohorts; \texttt{VQC Cyclic (CNOT)} shown in panels~\textbf{c}--\textbf{e}). (\textbf{b})~February--March 2022 halt quarantine ($320$ records, $13$ acute days, $\lVert\mathbf{x}\rVert_2 \le 74.4$) and post-reopen $R^2_{\text{OOS}}$ against the pre-crisis training average on unclipped and $\pm 5\sigma$-winsorized acute inputs, with acute unclipped $\max\,\lvert\hat{y}\rvert$ on the right. (\textbf{c})~$15\text{ bps}$ net active $\mathrm{IR}_{\text{LO}}$ ($95\%$ block bootstrap CI) across ten main models. (\textbf{d})~Break-even one-way costs ($c^* = 18.4$--$40.8\text{ bps}$) against the $15\text{ bps}$ baseline. (\textbf{e})~Deflated Sharpe Ratios ($\mathrm{DSR}_{\text{LO}}$ at $K_{\text{trials}} = 40$) against the $0.95$ threshold (global active long-only maximum $0.719$ for all $70$ evaluated series on \texttt{QCML Real}, $\lambda_E = 0$; Table~\ref{tab:economics_ood}A).}
  \label{fig:economics}
\end{figure*}

In Table~\ref{tab:economics_ood} and Fig.~\ref{fig:economics}, forecasts are translated into weekly MOEX portfolios with a one-day execution lag, $38$ masked short-ban rebalances, $500\text{ bps/yr}$ short-borrow costs and $15\text{ bps}$ flat or $10/30\text{ bps}$ tiered transaction costs.

\begin{table*}[t]
\centering
\caption{Causal market frictions of MOEX portfolios (\texttt{M+N}, $151$ weeks, one-day lag, $38$ masked short-ban rebalances, $500\text{ bps/yr}$ borrowing) and unclipped crisis OOD boundaries for February--March 2022 ($320$ rows / $13$ acute days; $R^2_{\text{OOS}}$ is evaluated against pre-crisis training averages). Panel~A contains flat $15\text{ bps}$ and tiered $10/30\text{ bps}$ active long-only information ratio ($\mathrm{IR}_{\text{LO}}$) over the equal-weight universe of stocks, single-seed average $\mathrm{IR}_{\text{LO}}$, dollar-neutral long--short Sharpe ($\mathrm{SR}_{\text{LS}}$), stationary block-bootstrap $95\%$ CIs, Deflated Sharpe Ratios ($\mathrm{DSR}_{\text{LO}} / \mathrm{DSR}_{\text{LS}}$ at $K_{\text{trials}}=40$ with strategy-specific autocorrelation-adjusted $T_{\text{eff}} \in [94.7, 130.6]$ for active LO in Panel~A; $125.5$ for \texttt{QCML Herm}) and standalone price-only long-only Sharpe above the average CBR key rate $13.34\%$ (Ex-CBR $\mathrm{SR}_{\text{LO}}$). Panel~B reports unclipped acute and post-reopen crisis $R^2_{\text{OOS}}$ and acute $\max\lvert\hat{y}\rvert$ across \texttt{M+N} and \texttt{MKT}.}
\label{tab:economics_ood}
\footnotesize
\setlength{\tabcolsep}{4pt}
\resizebox{\textwidth}{!}{%
\begin{tabular}{@{}lcccccc@{}}
\toprule
\multicolumn{7}{@{}l}{\textit{Panel A: Causal \texttt{M+N} portfolio performance ($15\text{ bps}$ flat and $10/30\text{ bps}$ tiered costs; $N_{\mathrm{seeds}}=10$ primary, $5$ for $\lambda_E=0$)}} \\
Model & $\mathrm{IR}_{\text{LO}}$ ($95\%$ CI) & Tiered $\mathrm{IR}_{\text{LO}}$ & \texttt{1s} $\mathrm{IR}_{\text{LO}}$ & $\mathrm{SR}_{\text{LS}}$ ($95\%$ CI) & $\mathrm{DSR}_{\text{LO}} / \mathrm{DSR}_{\text{LS}}$ & Ex-CBR $\mathrm{SR}_{\text{LO}}$ \\
\midrule
\texttt{Mom21-LowVol} & $+0.488$ ($[-0.58, +1.52]$) & $+0.411$ & $+0.488$ & $+0.069$ ($[-0.94, +1.08]$) & $0.087 / 0.018$ & $-0.055$ \\
\texttt{HistGBDT} & $+1.072$ ($[+0.15, +1.89]$) & $+0.828$ & $+1.072$ & $+0.629$ ($[-0.27, +1.55]$) & $0.519 / 0.172$ & $+0.207$ \\
\texttt{LightGBM} & $+0.969$ ($[+0.01, +1.85]$) & $+0.722$ & $+0.969$ & $+0.775$ ($[-0.15, +1.69]$) & $0.320 / 0.215$ & $+0.154$ \\
\texttt{MLP-tanh} & $+0.741$ ($[-0.18, +1.61]$) & $+0.500$ & $+0.467$ & $+0.537$ ($[-0.32, +1.49]$) & $0.210 / 0.142$ & $+0.065$ \\
\texttt{FiLM-MLP (iso)} & $+0.684$ ($[-0.13, +1.46]$) & $+0.444$ & $+0.301$ & $-0.170$ ($[-1.00, +0.65]$) & $0.223 / 0.006$ & $+0.018$ \\
\texttt{VQC Cyclic (CNOT)} & $+1.214$ ($[+0.34, +1.98]$) & $+0.998$ & $+0.933$ & $+1.204$ ($[+0.34, +2.08]$) & $0.637 / 0.620$ & $+0.250$ \\
\texttt{QCML Herm} ($d=8$) & $+1.026$ ($[+0.09, +1.83]$) & $+0.822$ & $+0.825$ & $+1.227$ ($[+0.33, +2.08]$) & $0.415 / 0.634$ & $+0.210$ \\
\texttt{QCML Real} ($d=11$) & $+1.114$ ($[+0.22, +1.91]$) & $+0.921$ & $+0.777$ & $+1.272$ ($[+0.44, +2.08]$) & $0.578 / 0.773$ & $+0.239$ \\
\texttt{QCML Diag} ($d=64$, $\beta^*$) & $+0.845$ ($[-0.10, +1.64]$) & $+0.656$ & $+0.580$ & $+0.514$ ($[-0.31, +1.39]$) & $0.311 / 0.168$ & $+0.138$ \\
\texttt{QCML Real} ($\lambda_E=0$) & $+1.303$ ($[+0.43, +2.05]$) & $+1.071$ & $+0.993$ & $+0.607$ ($[-0.23, +1.50]$) & $0.719 / 0.200$ & $+0.316$ \\
\texttt{50/50 Blend} (\texttt{MomLV}+\texttt{Herm}) & $+0.978$ ($[-0.04, +1.88]$) & $+0.837$ & $+0.978$ & $+0.874$ ($[-0.06, +1.81]$) & $0.311 / 0.320$ & $+0.202$ \\
\midrule
\multicolumn{7}{@{}l}{\textit{Panel B: Unclipped February--March 2022 crisis shock bounds ($320$ rows / $13$ acute days vs.\ pre-crisis training mean)}} \\
Model & \texttt{M+N} Acute $R^2_{\text{OOS}}$ & \texttt{M+N} Reopen $R^2_{\text{OOS}}$ & \texttt{M+N} Acute $\max\lvert\hat{y}\rvert$ & \texttt{MKT} Acute $R^2_{\text{OOS}}$ & \texttt{MKT} Reopen $R^2_{\text{OOS}}$ & \texttt{MKT} Acute $\max\lvert\hat{y}\rvert$ \\
\midrule
\texttt{QCML Herm} ($d=8$) & $-0.0037$ & $+0.0035$ & $0.0039$ & $-0.0033$ & $+0.0025$ & $0.0037$ \\
\texttt{QCML Real} ($d=11$) & $-0.0048$ & $+0.0054$ & $0.0058$ & $-0.0048$ & $+0.0012$ & $0.0050$ \\
\texttt{QCML Diag} ($d=64$, $\beta=2$) & $+0.0001$ & $+0.0014$ & $0.0023$ & $-0.0023$ & $-0.0008$ & $0.0040$ \\
\texttt{VQC Affine (CNOT)} & $-0.0085$ & $-0.0019$ & $0.0088$ & $-0.0042$ & $-0.0056$ & $0.0116$ \\
\texttt{FiLM-MLP (iso)} & $+0.0041$ & $-0.0042$ & $0.0075$ & $+0.0085$ & $-0.0088$ & $0.0072$ \\
\texttt{Ridge} & $-0.0208$ & $+0.0067$ & $0.0169$ & $-0.0208$ & $+0.0067$ & $0.0169$ \\
\texttt{MLP-tanh} & $-0.0426$ & $-0.0029$ & $0.0209$ & $+0.0168$ & $-0.0623$ & $0.0256$ \\
\texttt{HistGBDT} & $-0.0571$ & $-0.0261$ & $0.0485$ & $-0.0139$ & $-0.0270$ & $0.0232$ \\
\texttt{MLP-ReLU} & $-0.0837$ & $-0.0011$ & $0.0730$ & $-0.2239$ & $-0.8095$ & $0.1046$ \\
\bottomrule
\end{tabular}%
}
\end{table*}

With $15\text{ bps}$ on \texttt{M+N} the ten-seed \texttt{QCML Herm} ensemble provides an active long-only Information Ratio of $\mathrm{IR}_{\text{LO}} = +1.026$ ($p_{\text{boot}} = 0.0265$, break-even costs $c^* = 37.7\text{ bps}$, turnover $\mathrm{TO}_{\text{ann}} = 28.0\times/\text{yr}$), close to \texttt{HistGBDT} ($+1.072$, $28.4\times/\text{yr}$), \texttt{QCML Real} ($+1.114$) and \texttt{VQC Cyclic (CNOT)} ($+1.214$). On dollar-neutral long--short spreads, the ten-seed \texttt{QCML Herm} ensemble reaches $\mathrm{SR}_{\text{LS}} = +1.227$ ($p_{\text{boot}} = 0.0090$; from $+1.500$ at zero borrowing costs to $+0.845$ at $1{,}200\text{ bps/yr}$), ahead of the deterministic \texttt{HistGBDT} ($+0.629$, $p_{\text{boot}} = 0.1855$) and \texttt{LightGBM} ($+0.775$, $p_{\text{boot}} = 0.1095$); however, at the single-seed level, deterministic trees outperform \texttt{QCML Herm} in $\mathrm{IR}_{\text{LO}}$ ($+1.072$ vs.\ $+0.825$) and match its $\mathrm{SR}_{\text{LS}}$ (\texttt{LightGBM} $+0.775$ vs.\ $+0.766$), so the portfolio gain of QCML stems mainly from cross-seed variance reduction. Although \texttt{Mom21-LowVol} leads the full-panel $\overline{\mathrm{RankIC}}$ at lower turnover ($18.5\times/\text{yr}$), it carries over poorly to the extreme quintile portfolios ($\mathrm{IR}_{\text{LO}} = +0.488$, $\mathrm{SR}_{\text{LS}} = +0.069$) due to increased tracking variance in the tails; an equal-weight $50/50$ blend with \texttt{QCML Herm} restores $\mathrm{IR}_{\text{LO}} = +0.978$ and $\mathrm{SR}_{\text{LS}} = +0.874$.

Taking into account multiple testing and the risk-free rate moderates claims of standalone trading alpha (Table~\ref{tab:economics_ood}A). Among $K_{\text{trials}} = 40$ effective trials, \texttt{QCML Herm} reaches $\mathrm{DSR}_{\text{LO}} = 0.415$ and $\mathrm{DSR}_{\text{LS}} = 0.634$. The highest values for all $70$ evaluated model--modality specifications reach only $0.719$ for the active long-only strategy (\texttt{QCML Real} at $\lambda_E = 0$, Table~\ref{tab:economics_ood}A; vs.\ $0.686$ in Table~\ref{tab:headline_oos}) and $0.773$ for long--short (\texttt{QCML Real}, $d = 11$), which is far below the threshold of $0.95$ \cite{bailey_2014,harvey_2016,shen_2026}. Standalone price-only Sharpe ratios above the average MOEX policy rate $13.34\%$ are also modest ($+0.210$ for \texttt{QCML Herm}; $+0.207$ for \texttt{HistGBDT}). They set a conservative lower bound because split-adjusted closing prices exclude cash dividends (masking ex-dividend dates leaves $\overline{\mathrm{RankIC}}$ unchanged at $0.0723 \to 0.0721$; Supplementary Material, Section~S3).

In the unclipped February--March 2022 halt window ($320$ stock-days across $13$ sessions; Table~\ref{tab:economics_ood}B, Fig.~\ref{fig:economics}b), Rayleigh--Ritz eigenvalue bounds strictly limit the forecasts of \texttt{QCML Herm} ($\max\lvert\hat{y}\rvert = 0.0039$, acute phase $R^2_{\text{OOS}} = -0.0037$), avoiding large extrapolations of unbounded \texttt{MLP-ReLU} ($-0.0837$ on \texttt{M+N}; $-0.8095$ after resumption of trading on \texttt{MKT}) and \texttt{HistGBDT} ($-0.0571$). Winsorization of extreme halt-period inputs to $\pm 5\sigma$ reduces acute losses on \texttt{M+N} for \texttt{MLP-ReLU} ($-0.0837 \to -0.0468$) and \texttt{Ridge} ($-0.0208 \to -0.0078$). By contrast, \texttt{HistGBDT} remains unchanged on the outer leaf of the tree ($-0.0571$). Spectrally bounded \texttt{QCML}, \texttt{VQC} and \texttt{FiLM-MLP} do not require winsorization. Because commuting \texttt{QCML Diag} ($\max\lvert\hat{y}\rvert = 0.0023$, acute phase $R^2_{\text{OOS}} = +0.0001$) and bounded-head \texttt{FiLM-MLP (iso)} ($+0.0041$) achieve the same crisis stability, this protection stems from bounded output layers rather than from non-commuting operators.

\subsection{Summary of Supplemental Controls and Robustness Audits}\label{subsec:robust_summary}

Additional experiments in the Supplementary Material (Sections~S1--S3, Fig.~S1 and Tables~S1--S2) test solver accuracy and robustness across five dimensions. On synthetic manifolds, Hermitian QCML recovers analytical ground states on the unsupervised circle $S^1$ ($E_0 = 0.251$ vs.\ theoretical $0.250$) and in supervised tests on $S^2$ and the $d = 3$ spin-1 nonlinear quadrupole ($R^2_{\text{OOS}} = 0.994$--$1.000$), outperforming diagonal models at fixed $d$. Expanding diagonal QCML to an equal parameter budget ($d = 9$, $36$ parameters) restores classical parity ($0.999$). On MOEX, ranking accuracy is concentrated in liquid \texttt{Top-30} blue chips (\texttt{QCML Real} $+0.0966$ and \texttt{QCML Herm} $+0.0881$ vs.\ \texttt{HistGBDT} $+0.0570$), shifting by no more than $\lvert\Delta\rvert \le 0.0008$ under synthetic distressed-delisting injections ($k \in \{3,5,7\}$ exits per fold) and maintaining relative model rankings on continuous pre-2020 listings (where all models rise uniformly, for example \texttt{QCML Herm} $+0.0794$). Anchored QCML and boosted trees remain stable in the 2024--2025 high-rate subperiod ($16\%$--$21\%$ CBR rate), on strictly purged $h = 15$-day horizons and on $126$-day rolling windows. When jointly forecasting four market targets from a common state, QCML lags behind linear and neural baselines in persistent volatility, but outperforms comparable neural heads in noisy return ranking and drawdown calibration.

\section{Discussion, Limitations, and Conclusion}\label{sec:discussion}

\subsection{Mechanistic Interpretation and Operator Controls}\label{subsec:audit_standards}

After removing scale and capacity distortions from the models, our results clarify the behavior of matrix-form QCML on noisy equity panels. Normalizing the Frobenius regularization by $1/d$ eliminates the $8\times$ shrinkage penalty, which otherwise weakens the $d = 64$ diagonal model. Under equal regularization, the classical commuting Gibbs--RBF replacement matches the Hermitian ($d = 8$) and real-symmetric ($d = 11$) versions of QCML in the out-of-sample ranking ($p_{\text{NW}} = 0.8532$ and $0.0917$). Ground state energy anchoring ($\lambda_E = 0.05$) penalizes the operator variance $\sigma^2(\psi_0)$, pushing the trained non-commutativity $39\%$ below random GUE matrices and halving the variance of initializations without temperature tuning. News placebo tests with date permutation yield comparable drops for the Hermitian QCML ($0.0184$) and its commuting Gibbs replacement ($0.0232$), leaving the static \texttt{RFF-KRR} model and boosted trees statistically unchanged. Since daily news indicators are market-wide, cross-sectional lift requires multiplicative gating of stocks and news. End-to-end learned prototypes and subspaces provide this conditioning, rather than static Fourier bases, complex Berry phases, or non-commuting operators. During the 2022 trading halt, commuting diagonal QCML and bounded-output classical networks both clamp crisis forecasts. Along with the entanglement null ($0/6$) in the 3-qubit VQCs, these findings extend the de-quantization results of quantum kernel methods \cite{huang_2021,bowles_2024,schnabel_2025,fellner_2026,shen_2026} directly to Hamiltonian embeddings.

The second result explains the apparent sensitivity of raw spectral gaps to market crises. Because the quadratic term $\frac{1}{2}\lVert\mathbf{x}\rVert_2^2 I_d$ and the linear relationship $-\lVert\mathbf{x}\rVert_2\sum_k \hat{x}_k A_k$ are built into $H(\mathbf{x})$, Weyl's inequality forces $E_0(\mathbf{x})$ and $\Delta E(\mathbf{x})$ to expand whenever market volatility inflates the norms of the variables---even for untrained random matrices. Dividing by the radial amplitude or adding lagged norms in the Granger regressions eliminates the early warning power of the spectral gap under any objective function tested.

Along with standard empirical-finance evaluation hygiene---such as purged walk-forward splits, execution lags, trading costs, and multiple testing adjustments \cite{bailey_2014,harvey_2016,gu_2020,lopez_lira_2025,shen_2026}---our theoretical analysis points to two operator-specific controls that directly address methodological gaps in recent financial studies on QML and QCML \cite{samson_2024,gong_2025,hammond_2026,bowles_2024,schnabel_2025}:
\begin{enumerate}
    \item \emph{Capacity- and scale-matched algebra swaps.} Compare complex Hermitian models with real-symmetric ($F_{jk} \equiv 0$) and commuting diagonal Gibbs--RBF counterparts ($[A_j, A_k] = 0$) under dimension-invariant $1/d$ regularization and matched validation budgets, along with separable circuit ablations, to distinguish non-commutative geometry from generic capacity or smoothness.
    \item \emph{Radial-norm and random-matrix spectral null models.} Compare spectral regime indicators with $\lVert\mathbf{x}\rVert_2^2$, \texttt{GARCH(1,1)}, untrained GUE random matrices \cite{laloux_1999,plerou_2002}, scale-free $\widetilde{\Delta E}$ and norm-controlled Granger regressions to avoid mistaking mechanical Weyl scaling for learned geometric early warning.
\end{enumerate}

\subsection{Limitations and Future Work}\label{subsec:limitations}

The study is limited by four factors. First, our primary non-commuting QCML models use zero-temperature ground states ($\beta = \infty$) with $\lambda_E = 0.05$ stabilization, while the commuting control selects finite $\beta^*_f$ for each fold. Comparison with finite-temperature Hermitian Gibbs states $\rho_\beta(\mathbf{x})$ on matched $\beta$ grids remains a natural next step.

Second, stand-alone Sharpe ratios above the $13.34\%$ CBR policy rate are calculated using split-adjusted closing prices excluding cash dividends. Omitting dividends uniformly reduces cash excess Sharpe ratios across all models without affecting market-neutral or benchmark-relative spreads ($\mathrm{IR}_{\text{LO}}$, $\mathrm{SR}_{\text{LS}}$), and masking ex-dividend dates preserves the cross-sectional ranking ($0.0723 \to 0.0721$; Supplementary Material, Section~S3). Total return series would clarify cash-hurdle comparisons. At the same time, single-seed tree baselines (\texttt{HistGBDT} $+1.072$ vs.\ single-seed \texttt{QCML Herm} $+0.825$) remind practitioners that ensembling provides part of the portfolio stability of QCML.

Third, although spectral-gap opening ($\Delta E$) does not pass norm control, the Candelori unit-sphere energy $E_0^{\text{sphere}}$ retains directional sensitivity to stress ($0.860$--$0.911$ contemporaneous AUROC, just below \texttt{GARCH(1,1)} at $0.930$), capturing directional clustering on $S^{p-1}$. A direct comparison of $E_0^{\text{sphere}}$ with spherical PCA and Mahalanobis cosine distance on unit-normalized returns deserves a separate study.

Fourth, all $O(d^3)$ eigensolvers and 3-qubit state vector VQCs are exact classical simulations, not noisy hardware runs \cite{cerezo_2021,li_2026,thakkar_2024}. Hardware-native sparse Hamiltonians \cite{candelori_2026} remain outside the scope of our work. Although survivorship tests and continuous pre-2020 subsamples confirm minimal bias on MOEX (Supplementary Material, Section~S3), replication of the test on multi-country intraday panels of total returns remains valuable.

\subsection{Conclusion}\label{subsec:conclusion}

On the 2020--2025 panel of MOEX stocks and lagged news, neither complex Berry curvature, nor non-commutative operators, nor circuit entanglement, nor spectral-gap opening delivers statistically reliable superiority over matched classical controls \cite{shen_2026}. These results indicate a narrow classical niche rather than a quantum advantage. Matrix-form QCML acts as an adaptive spectrally bounded classical kernel: it integrates lagged macro news as well as the commuting Gibbs--RBF replacement, halves the spread of initializations under energy anchoring, achieves competitive information ratios on blue chips (below the $\mathrm{DSR} > 0.95$ multiple testing threshold) and curbs extreme forecasts at market halts. Treating Hamiltonian embeddings as regularized classical operators, tested against commuting substitutions and random-matrix null models, provides a transparent basis for quantum-inspired econophysics.


\FloatBarrier
\appendix
\renewcommand{\appendixname}{Appendix}
\renewcommand{\thesection}{\appendixname~\Alph{section}}
\makeatletter
\@addtoreset{table}{section}
\@addtoreset{equation}{section}
\makeatother
\renewcommand{\thetable}{\Alph{section}.\arabic{table}}
\renewcommand{\theequation}{\Alph{section}.\arabic{equation}}
\renewcommand{\thefigure}{A\arabic{figure}}
\renewcommand{\theHsection}{app.\Alph{section}}
\renewcommand{\theHfigure}{app.\arabic{figure}}
\renewcommand{\theHtable}{app.\Alph{section}.\arabic{table}}
\renewcommand{\theHequation}{app.\Alph{section}.\arabic{equation}}
\setcounter{figure}{0}
\section{Analytical Perturbation Gradients, Resolvent Stability, and Notation Bridge}\label{app:derivations}

\subsection{Exact Hellmann--Feynman and Rayleigh--Schr\"odinger Gradients}\label{app:gradients}

When the real vector space of Hermitian matrices of size $d \times d$ is endowed with the Frobenius inner product $\langle X, Y\rangle_F = \operatorname{Tr}(XY)$, any smooth real loss function $f$ has a unique Hermitian gradient $\nabla_O f$ satisfying $df = \operatorname{Tr}(\nabla_O f\, dO)$. Since the scalar shift $\frac{1}{2}\lVert\mathbf{x}\rVert_2^2 I_d$ in Eq.~\eqref{eq:hamiltonian} does not depend on $\mathcal{A}$, perturbation of the matrix $A_k$ gives the differential through the anticommutator
\begin{equation}\label{eq:app_dH}
dH(\mathbf{x}) = d\tilde{H}(\mathbf{x}) = \frac{1}{2}\bigl\{A_k - x_k I_d,\; dA_k\bigr\},
\end{equation}
where $\{X, Y\} = XY + YX$ preserves Hermiticity even when $[A_k, dA_k] \neq 0$.

For a non-degenerate ground state ($\Delta E(\mathbf{x}) = E_1(\mathbf{x}) - E_0(\mathbf{x}) > 0$) with projector $\rho_0(\mathbf{x}) = \lvert\psi_0(\mathbf{x})\rangle\langle\psi_0(\mathbf{x})\rvert$, differentiating $(H - E_0 I_d)\lvert\psi_0\rangle = 0$ gives $(H - E_0 I_d)\lvert d\psi_0\rangle = -(dH - dE_0 I_d)\lvert\psi_0\rangle$. Left multiplication by $\langle\psi_0\vert$ leads to the standard Hellmann--Feynman derivatives \cite{feynman_1939}:
\begin{equation}\label{eq:app_hf}
\begin{aligned}
\nabla_{A_k} E_0(\mathbf{x}) &= \frac{1}{2}\bigl\{A_k - x_k I_d,\; \rho_0(\mathbf{x})\bigr\}, \\
\frac{\partial E_0(\mathbf{x})}{\partial x_k} &= x_k - \bar{x}_k(\psi_0(\mathbf{x})).
\end{aligned}
\end{equation}
Supervised learning also requires differentiation of $\rho_0(\mathbf{x})$ within $\hat{y}_m(\mathbf{x}) = \operatorname{Tr}(\rho_0(\mathbf{x}) B_m)$. In the parallel transport gauge $\langle\psi_0\vert d\psi_0\rangle = 0$, the projection onto the excited subspace $Q_0(\mathbf{x}) = I_d - \rho_0(\mathbf{x})$ inverts the operator $(H - E_0 I_d)$ via the reduced resolvent $R_0(\mathbf{x})$:
\begin{equation}\label{eq:app_rs}
\begin{aligned}
\lvert d\psi_0(\mathbf{x})\rangle &= -R_0(\mathbf{x})\, d\tilde{H}(\mathbf{x})\,\lvert\psi_0(\mathbf{x})\rangle, \\
R_0(\mathbf{x}) &= \sum_{\alpha=1}^{d-1}\frac{1}{\Delta_\alpha(\mathbf{x})}\lvert\psi_\alpha(\mathbf{x})\rangle\langle\psi_\alpha(\mathbf{x})\rvert,
\end{aligned}
\end{equation}
with $\Delta_\alpha(\mathbf{x}) = E_\alpha(\mathbf{x}) - E_0(\mathbf{x}) > 0$. Here $\bar{B}(\mathbf{x}) = \sum_{m=1}^{q}(\partial \ell / \partial \hat{y}_m) B_m$ denotes the sensitivity of the supervised readout ($\bar{B} = 2(\hat{y} - y)B$ for scalar MSE), and substituting Eq.~\eqref{eq:app_rs} into $d\ell = \operatorname{Tr}(\bar{B}\,d\rho_0)$ and adding $\lambda_E dE_0$ gives the gauge-invariant gradient with respect to $\tilde{H}(\mathbf{x})$:
\begin{equation}\label{eq:app_gradH}
\begin{aligned}
G_{\tilde{H}}(\mathbf{x}) &\equiv \nabla_{\tilde{H}}\bigl(\ell + \lambda_E E_0\bigr) \\
&= -\bigl(R_0(\mathbf{x})\bar{B}(\mathbf{x})\rho_0(\mathbf{x}) + \rho_0(\mathbf{x})\bar{B}(\mathbf{x})R_0(\mathbf{x})\bigr) \\
&\quad + \lambda_E\,\rho_0(\mathbf{x}).
\end{aligned}
\end{equation}
Back-substituting Eq.~\eqref{eq:app_gradH} into Eq.~\eqref{eq:app_dH} automatically preserves Hermitian (or real-symmetric) structure. On a minibatch of size $M$, averaging gives the exact matrix updates for Eq.~\eqref{eq:loss}:
\begin{equation}\label{eq:app_gradAB}
\begin{aligned}
\nabla_{B_m}\mathcal{L} &= \frac{1}{M}\sum_{s=1}^{M}\frac{\partial \ell_s}{\partial \hat{y}_{s,m}}\,\rho_0(\mathbf{x}_s) + \frac{2\lambda_{\text{reg}, B_m}}{d}B_m, \\
\nabla_{A_k}\mathcal{L} &= \frac{1}{2M}\bigl\{A_k, \bar{G}\bigr\} - \frac{1}{M}S_k + \frac{2\lambda_{\text{reg}}}{d}A_k,
\end{aligned}
\end{equation}
where $\bar{G} = \sum_{s=1}^{M} G_{\tilde{H}}(\mathbf{x}_s)$ and $S_k = \sum_{s=1}^{M} x_{s,k} G_{\tilde{H}}(\mathbf{x}_s)$ are accumulated once per batch in $O(M(p+q)d^2 + p d^3)$ operations. The variance-weighted functional $\mathcal{L}_{\text{Cand}}(\mathcal{A})$ is derived identically by choosing $\bar{B}(\mathbf{x}) = \sum_{k=1}^{p}\bigl(2[(1 - w_{\text{var}})\bar{x}_k - x_k]A_k + w_{\text{var}} A_k^2\bigr)$, along with the direct operator derivative when $\rho_0$ is fixed. From $\partial H/\partial x_k = -(A_k - x_k I_d)$, the input Jacobian and the quantum geometric tensor \cite{zanardi_2007,candelori_2025,hammond_2026} also follow:
\begin{equation}\label{eq:app_jq}
\begin{aligned}
J_{mk}(\mathbf{x}) &= \frac{\partial \hat{y}_m}{\partial x_k} = 2\operatorname{Re}\langle\psi_0(\mathbf{x})\vert A_k R_0(\mathbf{x}) B_m\vert\psi_0(\mathbf{x})\rangle, \\
Q_{jk}(\mathbf{x}) &= \langle\psi_0(\mathbf{x})\vert A_j R_0(\mathbf{x})^2 A_k\vert\psi_0(\mathbf{x})\rangle \\
&= g_{jk}(\mathbf{x}) - \frac{\mathrm{i}}{2}F_{jk}(\mathbf{x}).
\end{aligned}
\end{equation}
Real-symmetric matrices have real eigenvectors, so the Berry curvature is zero ($F_{jk}(\mathbf{x}) \equiv 0$). At $\beta = \infty$, commuting diagonal matrices satisfy $R_0 A_k\lvert\psi_0\rangle = 0$, which gives $J_{mk}(\mathbf{x}) \equiv 0$ almost everywhere (piecewise-constant Voronoi partition).

\subsection{Tikhonov Resolvent Damping and Finite-Temperature Limit}\label{app:resolvent}

Since the unregularized resolvent $\lVert R_0(\mathbf{x})\rVert_2 = 1/\Delta E(\mathbf{x})$ diverges near avoided level crossings ($\Delta E(\mathbf{x}) \to 0$), we replace $\Delta_\alpha^{-1}$ in Eq.~\eqref{eq:app_rs} with the Tikhonov-damped resolvent:
\begin{equation}\label{eq:app_tikhonov}
\begin{aligned}
R_0^\varepsilon(\mathbf{x}) &= \sum_{\alpha=1}^{d-1}\frac{\Delta_\alpha(\mathbf{x})}{\Delta_\alpha(\mathbf{x})^2 + \varepsilon_{\text{gap}}^2}\,\lvert\psi_\alpha(\mathbf{x})\rangle\langle\psi_\alpha(\mathbf{x})\rvert, \\
\varepsilon_{\text{gap}} &= 10^{-5}.
\end{aligned}
\end{equation}
Because $\max_{u \ge 0} u/(u^2 + \varepsilon_{\text{gap}}^2) = (2\varepsilon_{\text{gap}})^{-1}$, $R_0^\varepsilon(\mathbf{x})$ is globally bounded while remaining cubic-accurate away from crossings:
\begin{equation}\label{eq:app_bound}
\begin{aligned}
\bigl\lVert R_0^\varepsilon(\mathbf{x})\bigr\rVert_2 &\le \frac{1}{2\varepsilon_{\text{gap}}}, \\
\bigl\lVert R_0^\varepsilon(\mathbf{x}) - R_0(\mathbf{x})\bigr\rVert_2 &= \frac{\varepsilon_{\text{gap}}^2}{\Delta E(\mathbf{x})\bigl(\Delta E(\mathbf{x})^2 + \varepsilon_{\text{gap}}^2\bigr)} \\
&= O\!\left(\frac{\varepsilon_{\text{gap}}^2}{\Delta E(\mathbf{x})^3}\right).
\end{aligned}
\end{equation}
Since $\lvert u\rangle = R_0^\varepsilon\bar{B}\lvert\psi_0\rangle$ is orthogonal to $\lvert\psi_0\rangle$, the supervised term in Eq.~\eqref{eq:app_gradH} satisfies $\lVert G_{\tilde{H}}(\mathbf{x}) - \lambda_E \rho_0(\mathbf{x})\rVert_F = \sqrt{2}\,\lVert u\rVert_2 \le \lVert\bar{B}(\mathbf{x})\rVert_2 / (\sqrt{2}\,\varepsilon_{\text{gap}})$ everywhere. For commuting diagonal matrices at finite $\beta$ with $B = \operatorname{diag}(b_0,\dots,b_{d-1})$, Eq.~\eqref{eq:gibbs_readout} reduces directly to the smooth classical softmax mixture $\hat{y}^{(\beta)}(\mathbf{x}) = \sum_{\alpha=0}^{d-1} w_\alpha(\mathbf{x}) b_\alpha$ without the need to diagonalize the matrices.

\subsection{Symmetry-Reduced Parameter Counts, $1/d$ Weight Decay, and GUE Baseline}\label{app:gauge}

Simultaneous unitary conjugation ($O \mapsto V O V^\dagger$) of all matrices in $\mathcal{A} \cup \mathcal{B}$ preserves eigenvalues, spectral gaps, commutator norms and forecasts ($\hat{y}_m(\mathbf{x})$). The redundant choice of the global basis corresponds to the group $\mathrm{PU}(d)$ of dimension $d^2 - 1$ for complex Hermitian matrices and $\mathrm{PO}(d)$ of dimension $d(d-1)/2$ for real-symmetric ones, and in the diagonal case it reduces to discrete permutations of coordinates. Quotienting gives the effective number of parameters:
\begin{equation}\label{eq:app_neff}
\begin{aligned}
N_{\text{eff}}^{\mathbb{C}} &= (p + q - 1)d^2 + 1, \\
N_{\text{eff}}^{\mathbb{R}} &= (p + q)\frac{d(d+1)}{2} - \frac{d(d-1)}{2}, \\
N_{\text{eff}}^{\text{diag}} &= (p + q)d.
\end{aligned}
\end{equation}
For \texttt{M+N} ($p = 18, q = 1$) Eq.~\eqref{eq:app_neff} gives $N_{\text{eff}} = 1{,}153$ ($d = 8$ Hermitian, $1{,}216$ raw), $1{,}199$ ($d = 11$ real-symmetric, $1{,}254$ raw) and $1{,}216$ ($d = 64$ diagonal). For \texttt{MKT} ($p_{\text{mkt}} = 12, q = 1$) the corresponding values are $769/832$, $803/858$ and $832/832$. The factor $1/d$ with Frobenius weight decay in Eq.~\eqref{eq:loss} equalizes the mean-square shrinkage of eigenvalues ($\frac{1}{d}\lVert O\rVert_F^2 = \frac{1}{d}\sum_{\alpha=0}^{d-1}\lambda_\alpha(O)^2$) across dimensions ($d \in \{2,4,8,11,16,64\}$).

For independent random matrices $\mathrm{GUE}(d)$ with zero mean and element covariance $\mathbb{E}[(A_j)_{ab}(A_j)_{cd}] = \sigma^2 \delta_{ad}\delta_{bc}$, the Wick contraction gives $\mathbb{E}\lVert A_j\rVert_F^2 = \sigma^2 d^2$, $\mathbb{E}\operatorname{Tr}(A_j^2 A_k^2) = \sigma^4 d^3$ and $\mathbb{E}\operatorname{Tr}(A_j A_k A_j A_k) = \sigma^4 d$, hence $\mathbb{E}\lVert[A_j, A_k]\rVert_F^2 = 2\sigma^4 d(d^2 - 1)$. Normalization by $\mathbb{E}(\lVert A_j\rVert_F^2\lVert A_k\rVert_F^2) = \sigma^4 d^4$ determines the RMS baseline $\kappa_{\text{nc}}^{\text{GUE}}(d) = \sqrt{\frac{2}{d}(1 - d^{-2})}$ in Eq.~\eqref{eq:kappa_nc}. At $d = 8$ it is equal to $0.4961$ and is in close agreement with the empirical average $0.4933$ from random draws.

\subsection{Dirac-to-Econometrics Notation Bridge}\label{app:glossary}

Consistent with the interdisciplinary practice of quantitative finance \cite{li_2026}, Table~\ref{tab:app_glossary} translates the quantum-mechanical notation from Section~\ref{sec:methods} into standard econometric terms.

\begin{table}[t]
  \centering
  \caption{Dirac-to-econometrics notation bridge for Section~\ref{sec:methods} (following \cite{li_2026}).}
  \label{tab:app_glossary}
  \footnotesize
  \setlength{\tabcolsep}{4pt}
  \begin{tabular*}{\columnwidth}{@{\extracolsep{\fill}}>{\raggedright\arraybackslash}p{0.34\columnwidth}>{\raggedright\arraybackslash}p{0.62\columnwidth}@{}}
    \toprule
    Quantum Object & Financial Econometrics Counterpart \\
    \midrule
    $\lvert\psi_0(\mathbf{x}_{i,t})\rangle$, $\rho_0(\mathbf{x}_{i,t})$ & Unit-norm latent state vector and rank-$1$ projector \\
    $A_k, B \in \mathbb{C}^{d\times d}$ & Learnable Hermitian feature and target weight matrices \\
    $H(\mathbf{x}_{i,t}), E_0(\mathbf{x}_{i,t})$ & Quadratic feature-mismatch matrix and minimum bias-plus-variance energy \\
    $\hat{y}_{i,t} = \operatorname{Tr}(\rho_\beta B)$ & Bounded nonlinear return forecast clamped to $[\lambda_{\min}(B),\lambda_{\max}(B)]$ \\
    $\Delta E(\mathbf{x}), \widetilde{\Delta E}(\mathbf{x})$ & Raw first spectral gap and norm-normalized directional gap \\
    $\kappa_{\text{nc}}(\mathcal{A}), Q_{\text{MW}}$ & Matrix non-commutativity ratio and circuit entanglement measure \\
    \bottomrule
  \end{tabular*}
\end{table}


\FloatBarrier
\section*{CRediT authorship contribution statement}
\textbf{\AuthorName:} Conceptualization, Methodology, Software, Validation, Formal analysis, Investigation, Data curation, Writing -- original draft, Writing -- review \& editing, Visualization.

\section*{Declaration of competing interest}
The author declares that they have no known competing financial interests or personal relationships that could have appeared to influence the work reported in this paper.

\section*{Funding}
This research did not receive any specific grant from funding agencies in the public, commercial, or not-for-profit sectors.

\section*{Declaration of generative AI and AI-assisted technologies in the manuscript preparation process}
During the preparation of this work the author used generative AI assistants (Google Gemini) in order to refactor simulation code, run automated reproducibility checks, and refine English phrasing. After using this tool/service, the author reviewed and edited the content as needed and takes full responsibility for the content of the published article.

\section*{Data, code, and supplementary material availability}
Historical stock prices and turnover are obtained from the public Moscow Exchange ISS interface, and the daily key rate is from the Central Bank of Russia. Reproduction code, chart scripts, summary tables and derived daily news aggregates are publicly archived on Zenodo (\url{https://doi.org/10.5281/zenodo.24525012}) and available at \url{https://github.com/alpha7en/QCML_Emerging_Market}. Copyrighted primary news texts from Lenta.ru are not redistributed. The online version of the article is accompanied by Supplementary Material (Sections~S1--S3, Fig.~S1 and Tables~S1--S2) detailing synthetic non-commuting positive controls, joint four-target forecasting and robustness checks for liquidity, ex-dividend dates, survivorship and horizon.


\FloatBarrier
\balance
{\sloppy\interlinepenalty=10000
\bibliographystyle{elsarticle-num}
\bibliography{references}
}

\end{document}